\documentclass[floatfix,aps,amsmath,amssymb,reprint,twocolumn,showpacs,showkeys,superscriptaddress]{revtex4-2}

\usepackage{amsmath,amsfonts,amsthm,amssymb,array,bm,braket,dcolumn,enumerate,flafter,float,graphics,latexsym,makeidx,multirow,natbib,verbatim,xfrac}
\usepackage[english]{babel}
\usepackage[mathscr]{euscript}
\usepackage[utf8]{inputenc}
\usepackage[usenames]{color}
\usepackage[pdftex,colorlinks=true,linkcolor=blue,urlcolor=blue,citecolor=blue]{hyperref}

\begin{document}

\title{Lingering times at resonance: The case of Sb-based tunneling devices}

\author{E. D. Guarin Castro}
\affiliation{Departamento de Física, Universidade Federal de São Carlos, 13565-905 São Carlos, SP, Brazil}
\email{ed.guarin.castro@gmail.com}

\author{A. Pfenning}
\affiliation{Technische Physik, Physikalisches Institut and Würzburg‐Dresden Cluster of Excellence ct.qmat, Am Hubland, D-97074 Würzburg, Germany}

\author{F. Hartmann}
\affiliation{Technische Physik, Physikalisches Institut and Würzburg‐Dresden Cluster of Excellence ct.qmat, Am Hubland, D-97074 Würzburg, Germany}

\author{A. Naranjo}
\affiliation{Departamento de Física, Universidade Federal de São Carlos, 13565-905 São Carlos, SP, Brazil}

\author{G. Knebl}
\affiliation{Technische Physik, Physikalisches Institut and Würzburg‐Dresden Cluster of Excellence ct.qmat, Am Hubland, D-97074 Würzburg, Germany}

\author{M. D. Teodoro}
\affiliation{Departamento de Física, Universidade Federal de São Carlos, 13565-905 São Carlos, SP, Brazil}

\author{G. E. Marques}
\affiliation{Departamento de Física, Universidade Federal de São Carlos, 13565-905 São Carlos, SP, Brazil}

\author{S. Höfling}
\affiliation{Technische Physik, Physikalisches Institut and Würzburg‐Dresden Cluster of Excellence ct.qmat, Am Hubland, D-97074 Würzburg, Germany}

\author{G. Bastard}
\affiliation{Département de Physique, Ecole Normale Supérieure de Paris (ENS/PSL), Université PSL (Paris Sciences and Letters), F75005, Paris, France}

\author{V. Lopez-Richard}
\affiliation{Departamento de Física, Universidade Federal de São Carlos, 13565-905 São Carlos, SP, Brazil}

\begin{abstract}
Concurrent natural time scales related to relaxation, recombination, trapping, and drifting processes rule the semiconductor heterostructures' response to external drives when charge carrier fluxes are induced. This paper highlights the role of stoichiometry not only for the quantitative tuning of the electron-hole dynamics but also for significant qualitative contrasts of time-resolved optical responses during the operation of resonant tunneling devices. Therefore, similar device architectures and different compositions have been compared to elucidate the correlation among structural parameters, radiative recombination processes, and electron-hole pair and minority carrier relaxation mechanisms. When these ingredients intermix with the electronic structure in Sb-based tunneling devices, it is proven possible to assess various time scales according to the intensity of the current flux, contrary to what has been observed in As-based tunneling devices with similar design and transport characteristics. These time scales are strongly affected not only by the filling process in the  $\Gamma$ and L states in Sb-based double-barrier quantum wells but also by the small separation between these states, compared to similar heterostructures based on As.

\end{abstract}

\maketitle

\section{Introduction}
Quantum tunneling heterostructures are envisaged for the development of optoelectronic devices operating at high speeds and high frequencies \cite{Soderstrom1990,Feiginov,Muttlak,Tavares2017,Zhang2018,Watson2019,Wang2022}, exploiting the intertwining between electrical currents and optical emissions produced during their operation. This correlation can be controlled by tuning the charge carrier dynamics determined by both the architecture and the materials used during the fabrication of the device. In this regard, resonant tunneling diodes (RTDs) offer a simple structure to explore and investigate charge carrier dynamics by combining quantum transport \cite{Buttiker1988,Jimenez1994,Jimenez1995,Lee2018,Brown2021}, electronic structure tuning by material parameters \cite{Sun,Muttlak2019,Guarin2020,Rebey2022}, thermalization and recombination processes \cite{Takeda2015,Zacharie2017,Suchet2017}, as well as excitation and relaxation mechanisms \cite{Harada1986, Tsuchiya1987,Kas1998,Teran_2009}. Understanding how these features intermix during the resonant tunneling of majority carriers through a double barrier structure (DBS) and how they affect the temporal evolution of the carrier dynamics in the quasi-bond states are the main objectives of this work.

For this purpose, we have decided to show the contrasting temporal evolution of the carrier dynamics in arsenic (As-) and antimony (Sb-) based DBSs and provide a unified description for seemingly divergent pictures. The analysis is performed by correlating the transport characteristics with the temporal evolution of the double-barrier quantum well (QW) optical response in both systems. This approach allows unveiling the dependence of the carrier dynamics on the flux of the majority carriers through the DBS, by means of the characterization of distinctive time scales observed at different current (voltage) conditions.

The results disclose an apparent counterintuitive relationship between transport and time-resolved optical measurements. While high conduction-band offsets in Sb-based DBSs induce longer escape times of majority carriers as compared with As-based DBSs, similar current densities in both systems may give the appearance of comparable time scales. However, they are not and we can demonstrate that the emergence of contrasting time scales in these systems results from the competition among various relaxation and recombination mechanisms inside the DBS, whose weights depend not only on the modulation of the charge carrier population with the current condition but also on particular structural parameters related to composition. In particular, filling-of-states processes, a small separation between $\Gamma$ and L bands inside the double-barrier QW, and non-resonant currents in Sb-based DBSs allow solving the seeming discrepancy between transport and time-resolved measurements, as well as explaining the stronger dependence of the carriers dynamics temporal evolution on the electrical current.

Our observations are supported by a model showing that, in Sb-based systems, the amount of current through the resonant channel enables the occurrence of a fast non-radiative relaxation process during out-of-resonance conditions and a slow recombination and minority-carriers relaxation dynamic at resonance. These findings provide clues to comprehend the temporal evolution of the carrier dynamics in quasi-two-dimensional quantized states, discerning the role of the minority carriers in the dynamics along with the influence of limiting factors related to the flux of the electrical current.

\section{Samples and Experimental Methods}
\begin{figure}
	\includegraphics{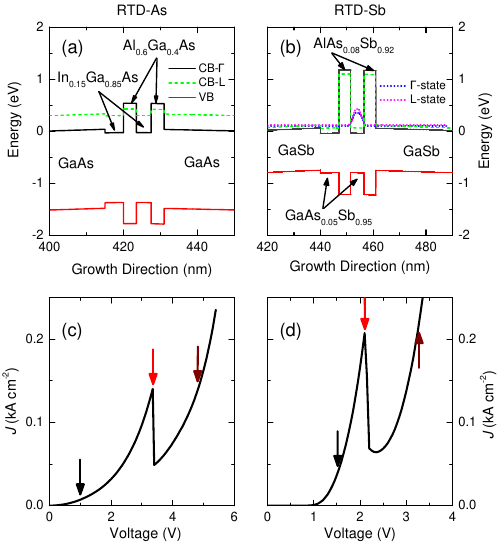}
	\caption{\label{Figure1} Simulated band profiles of the DBS for (a) RTD-As and (b) RTD-Sb, showing the conduction band minimum at the $\Gamma$- (black) and L-points (green), and the valence band maximum (red). Panel (b) also shows the simulated wavefunctions inside the double-barrier QW of RTD-Sb, for the first levels of the $\Gamma$ (blue) and L (pink) quasi-bond states. Current density as a function of voltage under illumination for (c) RTD-As and (d) RTD-Sb. Arrows point to the three different voltages where transient responses in Fig.~\ref{Figure2} are presented.}
\end{figure}

Two $n$-type RTDs were employed for this research, labeled in what follows as RTD-As and RTD-Sb. Both were prepared via molecular beam epitaxy with pseudomorphically grown ternary emitter prewells. The As-based RTD, used as a reference sample, was grown on a Si $n$-doped GaAs substrate, including an In$_{0.15}$Ga$_{0.85}$As emitter prewell and a double-barrier QW with thicknesses of 5 and 4 nm, respectively. The QW is sandwiched by two 3.5-nm thick Al$_{0.6}$Ga$_{0.4}$As barriers, and the DBS is surrounded by two 20-nm thick undoped GaAs spacer layers, as displayed in Fig.~\ref{Figure1} (a) by the simulated band profile of the conduction band (CB) at the $\Gamma$ (solid black line) and L (dashed green line) minimums, and the valence band (VB) maximum (solid red line) \cite{Birner2007}. CB barriers at the $\Gamma$ minimum with a height of around 0.5~eV are expected in this kind of DBSs. Moreover, the separation between the $\Gamma$ and L minima inside the double barrier QW is $\Delta E_\text{$\Gamma$-L} \approx 0.33$~eV, which prevents electrons from occupying L states. A 300~nm thick high-bandgap Al$_{0.2}$Ga$_{0.8}$As optical window was deposited on top of the heterostructure to avoid optical loss by absorption and guarantee optical access for infrared wavelengths. The details of the RTD-As structure are outlined in Ref.~\cite{Cardozo2021} where it was labeled as `S-InGaAs'. 

In turn, RTD-Sb is an Sb-based RTD grown on a Te $n$-doped GaSb(100) substrate. Its DBS is composed of an emitter prewell and a double-barrier QW of GaAs$_{0.15}$Sb$_{0.85}$, with thicknesses of 5 and 7 nm, respectively. The QW is surrounded by two 4.5-nm thick AlAs$_{0.08}$Sb$_{0.92}$ barriers and the DBS is enclosed by two 20-nm thick undoped GaSb spacer layers, as represented in Fig.~\ref{Figure1} (b) \cite{Birner2007}. Here, CB barriers are higher than those in RTD-As, with heights of around 1.20~eV at the $\Gamma$ minimum. In addition, the $\Gamma$-L energy separation inside the double-barrier QW is $\Delta E_\text{$\Gamma$-L} \approx 0.10$~eV, which is lower as compared with RTD-As. At the top of the structure a 220~nm thick high-bandgap Al$_{0.30}$Ga$_{0.70}$As$_{0.03}$Sb$_{0.97}$ optical window was deposited, also to reduce optical losses and to favor infrared optical access. The heterostructure layout of RTD-Sb is fully described in Ref.~\cite{Pfenning2017-1} where it was labeled as `RTD 3'.

Both samples were cooled down to a nominal temperature of $T=4$~K in an ultra-low vibration cryostat (Attocube AttoDRY1000), associated with a homemade confocal microscope and a SourceMeter (Keithley 2400), to study their transport characteristics and the temporal evolution of the optical response emitted from their QWs at different applied voltages. Two excitation lasers (PicoQuant LDH Series) were employed for the optical excitation: one laser with emission energy $\hbar\omega=1.70 \text{ eV}$ and an optical power density of $2.29 \text{ kW}/\text{cm}^{2}$ to excite RTD-As, and another laser with energy $\hbar\omega=1.15 \text{ eV}$ and an optical power density of $12.5\text{ kW}/\text{cm}^{2}$, used to excite RTD-Sb. Both lasers were operated in continuous-wave and pulsed modes to characterize the emission spectra and the temporal evolution via photoluminescence (PL) and time-resolved PL spectroscopies, respectively. During PL measurements, the optical responses of RTD-As
and RTD-Sb observed at different bias voltages were dispersed by a 75 and 50~cm spectrometer (Andor Shamrock), respectively. Then, the signals were detected by a high-speed Si charge-coupled device (Andor iDus 420) and a high-resolution InGaAs diode array (Andor DU491A) for RTD-As and RTD-Sb, respectively. During time-resolved PL measurements, both lasers operated at a repetition rate of $80\text{ MHz}$, with a pulse duration of around $100\text{ ps}$, while the transient responses were detected by an infrared photomultiplier tube (PicoQuant Si PMT Hybrid and Hamamatsu InGaAs/InP H10330B-75, for RTD-As and RTD-Sb measurements, respectively) coupled to a time-correlated single-photon counting electronics (PicoQuant PicoHarp300).

\section{Transport and Time-resolved characteristics}

\begin{figure}
	\includegraphics{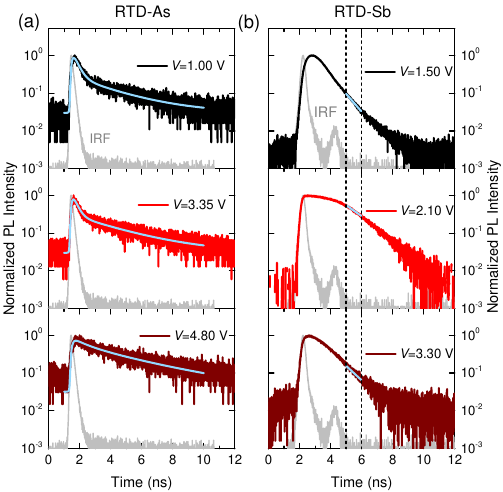}
	\caption{\label{Figure2} Normalized photoluminescence intensity of the double-barrier QW transient response in (a) RTD-As and (b) RTD-Sb, before (top), on- (middle), and after (bottom) resonance. Light-blue lines on the transient responses represent the results of the fitting procedure performed to extract the effective lifetimes. Gray lines in the background show the IRF.}
\end{figure}

The current-density, $J(V)$, characteristics observed under illumination for RTD-As and RTD-Sb are depicted in Figs.~\ref{Figure1} (c) and (d), showing resonances for majority carriers transport at 3.35 and 2.10~V, with current densities peaks of 0.14 and 0.21~kA cm$^{-2}$, and peak-to-valley current ratios of 2.9 and 3.2, respectively. Despite CB barriers in RTD-As being lower and thinner than the barriers in RTD-Sb, the current densities and peak-to-valley current ratios are comparable in both systems, which in principle implies similar escape times for electrons inside the double-barrier QWs.

In order to verify this assertion, time-resolved PL measurements were carried out in both systems. The time-resolved spectra for RTD-As and RTD-Sb at different applied voltages are shown in Figs.~\ref{Figure2} (a) and (b), respectively. The decay curves before (top panels), at (medium panels), and after resonance (bottom panels) are presented for the voltages indicated by vertical arrows in the $J(V)$ characteristics in Figs.~\ref{Figure1} (c) and (d). For RTD-As,  the curves are presented at $V=1.00$~V (black), $V=3.35$~V (red), and $V=4.80$~V (dark red). For RTD-Sb the responses are depicted at $V=1.50$~V (black), $V=2.10$~V (red), and $V=3.30$~V (dark red). In both cases, the corresponding voltages after resonance were chosen to produce the same current condition as on-resonance conditions.

All transient responses were detected at 1.55 and 0.93~eV for RTD-As and RTD-Sb, respectively, which correspond to the energies of maximum QW PL emission produced by the radiative recombination between the first confined levels of the CB $\Gamma$ minimum and the VB maximum in the double-barrier QWs. The PL intensities of the transient responses have been normalized and the Instrument Response Function (IRF) for each setup is also presented for reference as gray background curves. The IRF was measured by the analysis of the back reflected laser from the sample surface.

In the case of RTD-Sb, the IRF shows an after-pulse at around 4.3 ns, unlike the IRF for RTD-As, as presented in Fig.~\ref{Figure2} (b). This after-pulse peak is produced at a high counting regime by the elastic backscattering of electrons at the first dynode of the photomultiplier tube~\cite{Hamamatsu2007}, which is induced by the high amplification gain and the high voltage ($\sim 800$~V) in between the photocathode of the photomultiplier tube. Yet, the contribution of this peak to the transient response is negligible since it represents only $0.01\%$ of the counts in the main peak and is out of the range used for the fitting procedures. 

The transient curves for RTD-As in Fig.~\ref{Figure2} (a) reveal a fast exponential decay followed by a slow one, for both before and on-resonance conditions. At higher voltages, only the slow decay prevails. In order to extract effective lifetimes from these decays, a standard reconvolution procedure~\cite{Picoquant2018} was performed to fit the RTD-As decay curves. This procedure allows avoiding the influence of the IRF on the fast decay. The results of the fitting procedures are shown as blue curves.

In contrast, the transient curves of RTD-Sb displayed in Fig.~\ref{Figure2} (b) show a shoulder-like emission at the peak of maximum intensity observed after 2~ns, which hampers the implementation of reconvolution models for the extraction of the effective lifetimes. However, during resonance conditions at $V=2.10$~V (medium panel), the slow intensity decay produces a plateau that extends up to $t\approx 5$~ns. This decay can be fitted by using a mono-exponential function of the form $\exp(-t / \tau_\text{eff})$, which gives an effective lifetime of $\tau_\text{eff}=6$~ns. This value is similar to the limit of radiative recombination reported for GaAsSb/GaAs QWs~\cite{Baranowski2012}. The same fitting procedure was performed for the decay spectra between 5 and 6~ns (framed within vertical dashed lines), as indicated by straight blue lines in Fig.~\ref{Figure2} (b). Thus, the influence of either the IRF, the plateau, or highly noisy regions at longer times was avoided.

\begin{figure}
	\includegraphics{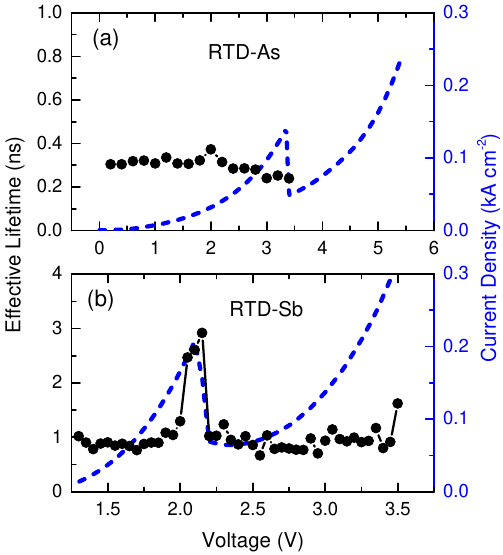}
	\caption{\label{Figure3} Effective lifetimes as a function of the applied voltage for (a) RTD-As and (b) RTD-Sb. The corresponding $J(V)$ characteristics have been plotted as blue dashed lines.}
\end{figure}

The effective lifetimes obtained by means of the fitting procedures just described are displayed in Figs.~\ref{Figure3} (a) and (b) as a function of the applied voltage for RTD-As and RTD-Sb, respectively. The corresponding $J(V)$ characteristics have also been plotted for reference (blue dashed lines). Note that, the effective lifetimes in RTD-As have a negligible voltage dependence, presenting values of $\sim0.3$ and $\sim3.0$~ns for the fast and slow (not presented here) decays, respectively. The slow decay can be attributed to an accumulation of photogenerated carriers at the prewell and their subsequent non-resonant tunneling~\cite{Kas1998,Romandic2000}. Thus, the fast decay can be associated with the carrier dynamics inside the double-barrier QW which can be affected by radiative and non-radiative recombination processes. In contrast, RTD-Sb shows a remarkable non-monotonic voltage dependence of the effective lifetime with the applied voltage. For out-of-resonance conditions, $\tau_\text{eff}$ shows a value of $\sim0.9$~ns  but at resonance, it increases with the current, peaking at $\sim3.0$~ns. This leads to a factor of $3$ when comparing the out-of-resonance effective lifetimes of both RTDs and a full-order factor in resonant conditions.

\section{Charge Carrier Dynamics in a DBS}
The difference between effective lifetimes in both samples can be initially explained by the nature of the radiative processes in each type of system. In bulk GaSb samples,~\cite{Levinshtein1996} radiative lifetimes are longer than in bulk GaAs samples,~\cite{Hooft1983} depending on the temperature and donor concentrations. As reported in Ref.~\cite{Bockelmann1993}, the radiative lifetime $\tau_0$ depends on the energy of the optical emission $\hbar\omega_0$ according to the expression $\tau_0 \propto (|\braket{\psi_\text{e} | \psi_\text{h}}|^2 E_p r \hbar\omega_0)^{-1}$, where $|\braket{\psi_\text{e} | \psi_\text{h}}|^2$ accounts for the overlapping between electron and hole wave functions, $E_p$ is the Kane energy, and $r$ is the refractive index in the material~\cite{Cesar2011}. As a consequence, higher emission energies imply shorter radiative lifetimes. Then, by taking the ratio between the double-barrier QW emission energies for both RTDs, one gets a factor of $1.55/0.93 \approx 1.7$ which is almost half the factor obtained when comparing their effective lifetimes. This result suggests that recombination processes alone are not enough to account for the discrepancy between effective lifetimes in both samples.

\subsection{Temporal evolution of carriers population and escaping times}
To understand the effective lifetime dependence on the current condition in both samples, a three-level rate equation model has been considered to study the carrier dynamics inside the DBSs. The model contemplates carrier interactions via radiative recombination as well as relaxation or escape processes. In this way, the time evolution of the carriers population in the double-barrier QWs, after their injection by resonant or non-resonant channels, can be described as
\begin{align}
\frac{dn_\text{E}}{dt} &= S_\text{e}  -\frac{n_\text{E}}{\tau_\text{e}} - \frac{n_\text{E}}{\tau_\text{T}} \left(1 - \frac{n}{N_\text{e}}\right), \label{eq_nE}\\
\frac{dn}{dt} &= \frac{n_\text{E}}{\tau_\text{T}} \left( 1 - \frac{n}{N_\text{e}}\right) - \frac{n p}{\tau_0} - \frac{n}{\tau_\text{e}}, \label{eq_n} \\
\frac{dp}{dt} &= S_\text{h} - \frac{n p}{\tau_0} - \frac{p}{\tau_\text{h}}, \label{eq_p}
\end{align}
where $n_\text{E}$ is the electron density at the emitter side while $n$ and $p$ are the electron and hole densities in the ground quasi-bond state of the double-barrier QWs. $S_i$ and $\tau_i$ are respectively the current source and lifetime for electrons and holes ($i=$e,h). $N_\text{e}$ is the electronic density of states, and $\tau_\text{T} \propto \mathcal{T}^{-1}$ is the electron tunneling time through the emitter barrier proportional to the transmission probability $\mathcal{T}$. In turn, $\tau_0$ represents the optical recombination time and the term $\left(1- n / N_\text{e}\right)$ refers to the saturation process in the double-barrier QWs caused by the finite density of states. This saturation can induce a plateau in the transient response at the beginning of the optical decay that has also been detected in Ref.~\cite{Baranowski2012}. 

\begin{figure}
	\includegraphics[width=8.6cm]{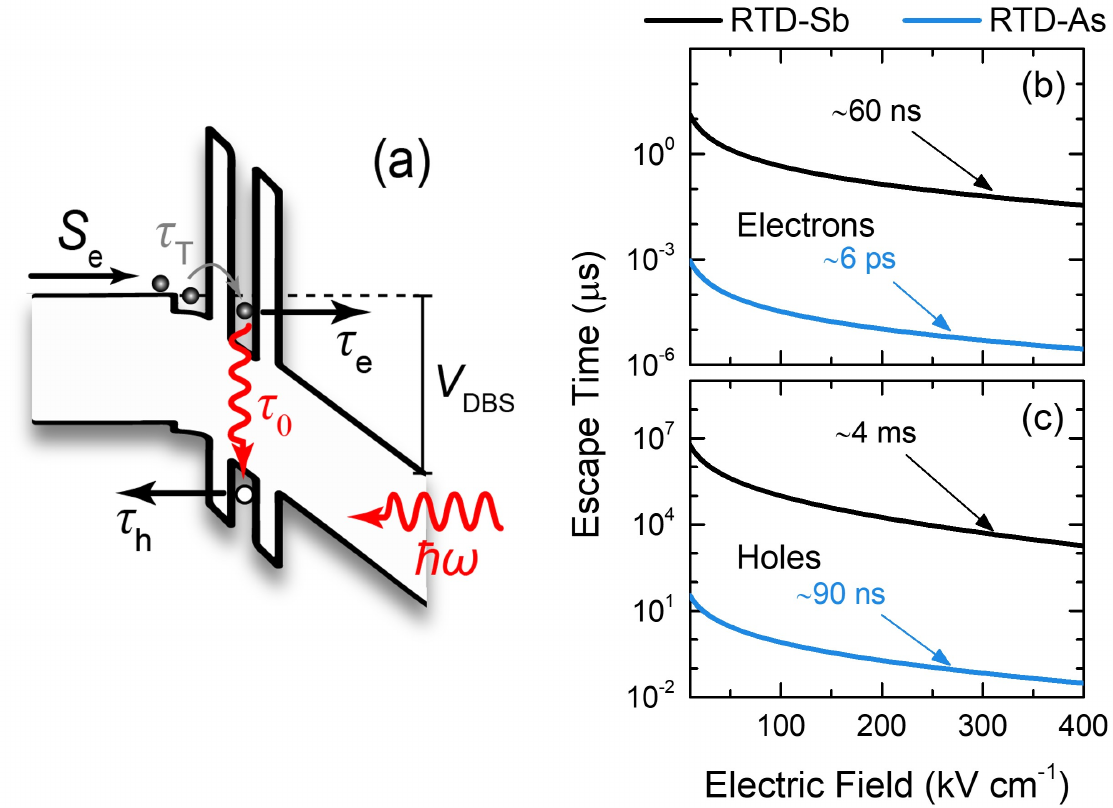}
	\caption{\label{Figure4} (a) Sketch of the carrier relaxation dynamics in the double-barrier QW after optical excitation. Calculated escape times for (b) electrons and (c) holes in the GaSb (black line) and GaAs (blue line) RTDs. In each case, arrows point to the obtained escape times at the resonance condition.}
\end{figure}

Figure~\ref{Figure4} (a) depicts the processes considered in the three-level model after excitation with a laser pulse of energy $\hbar\omega$, and when the DBS is under an applied bias voltage which produces a voltage drop $V_\text{DBS}$. The model also assumes a constant source of electrons $S_\text{e}$ responsible for the injection of majority carriers from the emitter side into the double-barrier QW. The injection of electrons also depends on the electron tunneling time $\tau_\text{T}$. At the resonance voltage, the injection rate increases due to the accumulation of electrons in the prewell which acts as a reservoir for electrons.

On the other hand, holes are efficiently transported through the DBS at the electron resonant condition, which hampers their accumulation at the right-hand side of the DBS in Fig.~\ref{Figure4} (a) but increases the probability of optical recombination inside the double-barrier QW. The absence of hole accumulation at the resonant electron condition is a consequence of the low barrier height in the valence band~\cite{Park1995,Coelho2004,Pfenning2016}. Consequently, accumulation of holes can only be detected at low voltages, as measured by the presence of a voltage shift in the J(V) characteristic under illumination for the same RTD-Sb~\cite{Guarin2021}, as well as for a p-type Sb-based RTD~\cite{Pfenning2018}.

Inside the double-barrier QW, carriers can radiatively recombine with a time scale $\tau_0$ or decay via escape or non-radiative recombination processes with different time scales $\tau_e$ and $\tau_h$ for electrons and holes, respectively. The latter, which for simplicity will henceforth be referred to as the escape times, can be estimated by considering a ping-pong model inside the QW. According to this model, a carrier oscillates classically inside the well with a period $\tau_\text{osc}(E)$~\cite{Bastard1991}, with $E$ being the mechanical energy. The semiclassical approximation of the escape frequency is then given by~\cite{Bastard1991,Larkin2009}
\begin{equation}\label{t_esc}
    \frac{1}{\tau_i} = \mathcal{T}_i(E) \frac{1}{\tau_{\text{osc},i}(E)},
\end{equation}
with $i=\text{e,h}$ and $\tau_{\text{osc},i}(E)$ as the period of the classical oscillations driven by an electrical force, $eF$, where $F$ is the local electric field. Then, the solution of the classical equation of motion gives as a result~\cite{bastard1990,Bastard1991}
\begin{equation}\label{t_osc}
    \tau_{\text{osc},i}(E) = 2 \sqrt{\frac{2m^*_i E_i}{(eF)^2}},
\end{equation}
where $m^*_i$ and $E_i$ are, respectively, the carriers' effective mass and confinement energy in the double-barrier QW. If a triangular well is considered due to the applied voltage, then $E_i$ is given by~\cite{Bastard1991}
\begin{equation}\label{E_conf}
    E_i = \zeta_1 eF L_\text{QW} \left[\frac{\hbar^2}{2m^*_i eF L^3_\text{QW}}\right]^{1/3},
\end{equation}
where $\zeta_1=-2.33$ corresponds to the first zero of the Airy Function ($\text{Ai}(\zeta_1)=0$), and $L_\text{QW}$ is the thickness of the double-barrier QW. In turn, the local electric field at the DBS can be calculated as $F=|V_\text{DBS}|/l_\text{DBS}$, with $V_\text{DBS}$ as a function of the applied voltage $V$, and defined by ~\cite{Guarin2021_phd}
\begin{equation}\label{VDBS}
   |V_{\text{DBS}}(V)| = -V_0 \left(1-\sqrt{1+\frac{2|V|}{V_0}}\right),
\end{equation}
with $V_0=el^2_{\text{DBS}} N_{D}^+/\varepsilon$, $l_\text{DBS}$ as an effective DBS length, $N_{D}^+$ as a constant 3D donor density, and $\varepsilon$ as the permittivity. For the calculations, effective permitivitties of $\varepsilon = 13 \varepsilon_0$ and $15 \varepsilon_0$, effective DBS length of $l_\text{DBS}=31$ and 41~nm, and nominal donor densities of $N^+_D = 2\times 10^{17}$~cm$^{-3}$ and $5\times 10^{17}$~cm$^{-3}$, were employed for RTD-As and RTD-Sb, respectively. Thus, electric fields of around 260 and 310~kV cm$^{-1}$ can be expected for RTD-AS and RTD-Sb, respectively, under resonance conditions.

By considering a thick enough rectangular barrier, the transmission in Eq.~\ref{t_esc} for different effective masses inside and outside the barrier was taken as
\begin{equation}\label{T_b}
    \mathcal{T}_i(E_i) = 16 \left(\Upsilon_i + \frac{1}{\Upsilon_i}\right)^{-2} \exp(-2 k_{\text{b},i} L_\text{b}),
\end{equation}
which is an adaptation of the expression for equal effective masses that can be found in Ref.~\onlinecite{Manasreh2005}. Here, $L_\text{b}$ is the barrier thickness, $k_{\text{b},i} = \sqrt{2 m^*_{\text{b},i} (U_{\text{b},i} - E_i)} / \hbar$, $\Upsilon_i = k_i m^*_{\text{b},i} / k_{\text{b},i} m^*_i$, and $k_i = \sqrt{2 m^*_i E_i} / \hbar$. Here, $m^*_{\text{b},i}$ and $m^*_{i}$ are the carrier effective masses in the barrier and QW, respectively, and $U_{\text{b},i}$ is the barrier height.

In our calculations, escape times were determined for electrons and holes in both samples by replacing Eqs.~\ref{t_osc}-\ref{T_b} into Eq.~\ref{t_esc} and taking the nominal values of the barrier and double-barrier QW thicknesses. For electrons $m^*_\text{e}=0.057m_0$ and 0.041$m_0$, $m^*_{\text{b,e}}=0.100m_0$ and 0.123$m_0$~\cite{Madelung2004,Vurgaftman2001}, and $U_{\text{b,e}}=0.55$ and 1.23~eV for RTD-As and RTD-Sb, respectively. In turn, for holes $m^*_\text{h}=0.495m_0$ and 0.405$m_0$, $m^*_{\text{b,h}}=0.600m_0$ and 0.919$m_0$~\cite{Madelung2004,Vurgaftman2001}, and $U_{\text{b,h}}=0.40$ and 0.47~eV for RTD-As and RTD-Sb, respectively. Barrier heights in both samples were determined by means of the calculated band profiles shown in Figs.~\ref{Figure1} (a) and (b).

Figures~\ref{Figure4} (b) and (c) display the calculated escape times from the double-barrier QW for electrons and holes, respectively, as a function of the local electric field in the DBS. Black and blue lines represent the escape times for RTD-As and RTD-Sb, respectively. Results suggest that the tunneling probability is reduced and consequently, the escape time increases in RTD-Sb due to its higher CB offsets as compared with RTD-As. The increase in the escape time for RTD-Sb is also a consequence of thicker barriers, as supported by calculations using Eq.~\ref{t_esc} when varying the barrier thickness (not shown here). The reduction of the tunneling rate for thicker barriers has been also observed in different RTDs, as reported in Refs.~\onlinecite{Tsuchiya1987,VanHoof1992_TRPL}. A lower tunneling rate, as expected for RTD-Sb, can lead to a reduction in the resonance current, contrary to the observed current-density characteristics shown in Figs.~\ref{Figure1} (c) and (d). However, these current densities are the result of the addition of coherent and incoherent contributions~\cite{Edson}. The former current is associated with resonant tunneling processes, while the latter corresponds to a sequential tunneling of carriers that lost their phase coherence and energy due to scattering processes. Consequently, incoherent currents in RTD-Sb can be higher than in RTD-As, thus making current densities comparable in both systems.

Escape times exhibited in Figs.~\ref{Figure4} (b) and (c) also show that electrons can escape from the double-barrier QW faster than holes, which can present slow relaxation dynamics in both systems. In RTD-Sb, electron escape times can be also of the order of nanoseconds, similar to the order of the optical recombination time obtained from the plateau of the transient curves. These results give an indication of the order of magnitude expected for carriers' escape times which is necessary to solve the set of Eqs.~\ref{eq_nE}-\ref{eq_p} and then, find the correlation between effective lifetimes and voltage conditions, as presented in the following sections.

\subsection{Transmission dependence of carriers population}
Based on the above considerations, the analysis of the carrier dynamics starts, for sake of simplicity, by considering optical recombination times longer than any other time-scale, $\tau_0 \to \infty$. This allows neglecting the second term on the right side of Eqs.~\ref{eq_n} and \ref{eq_p}. In addition, since electrons are majority carriers inside the double-barrier QW, then changes produced in the number of holes injected by a constant source $S_\text{h}$ are also negligible and we can set $S_\text{h}/S_\text{e} \to 0$. Consequently, the solutions for the equation system depend on the initial number of electrons and holes produced by the laser pulse and considered as $n_\text{E}(0)=n^0_\text{E}+\Delta n_\text{E}$, $n(0)=n_0$, and $p(0)=S_\text{h} \tau_\text{h} + \Delta p_0$ for electrons at the emitter side, and electrons and holes inside the double-barrier QW, respectively. Here $\Delta n_\text{E}$ and $\Delta p_0$ are variations in the initial carrier populations produced by photocreation processes. Then, by assuming the stationary condition $\Delta n_\text{E}=\Delta p_0=0$, one obtains
\begin{equation}\label{ne0}
    n^0_\text{E} = \frac{S_\text{e}}{\frac{1}{\tau_\text{e}} + \frac{1}{\tau_\text{T}}\left(1-\frac{n_0}{N_\text{e}}\right)},
\end{equation}
and $n_0 = (-b+\sqrt(b^2 + 4a))/2a$ where $a=-1/(S_\text{e}N_\text{e}\tau_\text{e})$ and $b=1/N_\text{e} + \tau_\text{T}/(S_\text{e}\tau^2_\text{e}) + 1/(S_\text{e}\tau_\text{e})$.

\begin{figure}
	\includegraphics{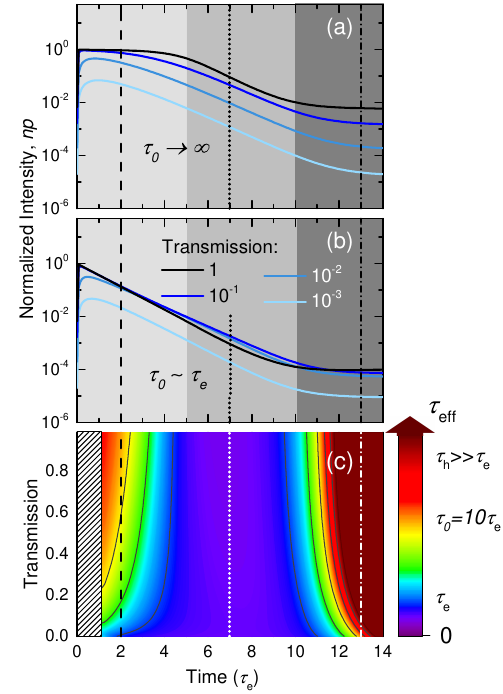}
	\caption{\label{Figure5} Simulated transient curves for (a) $\tau_0 \to \infty$ and (b) $\tau_0 \sim \tau_\text{e}$, and for different values of the transmission coefficient (blue lines). Short, intermediate, and long temporal ranges are highlighted with light-gray, gray, and dark-gray shadow regions, respectively. (c) Color-gradient map showing the calculated effective lifetime, $\tau_\text{eff}$, as a function of the transmission and time in units of $\tau_\text{e}$ for $\tau_\text{h} > \tau_0 > \tau_\text{e}$. Peaks of maximum intensity in the decay curves hamper the extraction of $\tau_\text{eff}$ for $t<\tau_\text{e}$.}
\end{figure}

Based on these initial conditions, Eqs.~\ref{eq_nE}-\ref{eq_p} were solved numerically for $n$ and $p$, by assuming $\tau_\text{h}/\tau_\text{e}=10^2$ and $S_\text{h}/S_\text{e}=0.1$. The optical recombination intensity was then calculated as the product between carrier populations, $np$, and plotted as a function of time as shown in Figs.~\ref{Figure5} (a) and (b) for $\tau_0 \to \infty$ and $\tau_0 \sim \tau_\text{e}$, respectively. Blue lines indicate transient responses at different transmission probabilities.

When $\tau_0 \to \infty$, the calculated response is similar to the transient curves observed for RTD-Sb. Under this condition, three time scales can be obtained at three different temporal ranges as indicated by shadow regions in Fig.~\ref{Figure5} (a): a slow decay at short ranges (light gray region), a fast decay at intermediate ranges (gray region), and another slow decay at long ranges (dark gray region). The former decay is enhanced when $\mathcal{T}\to 1$ (black line) or $\Delta n_\text{E}\to \infty$. This decay is a consequence of the filling process in the quasi-bond states of the double-barrier QW.

Taking into account that the recombination time extracted from the plateau displayed in Fig.~\ref{Figure2} (b) at the resonance condition for RTD-Sb is 6~ns, then it is expected that the escape time for electrons in this DBS must be in the sub-nanosecond scale. However, this estimation differs from the value of $\sim60$~ns presented in Fig.~\ref{Figure4} (b). This discrepancy can be ascribed to the contribution of faster escape channels from the $\Gamma$ ground state. According to the band profile simulation of RTD-Sb presented in Fig.~\ref{Figure1} (b) \cite{Birner2007}, the separation between $\Gamma$ (blue dotted line) and $L$ (pink dotted line) ground states at the double-barrier QW is just $\Delta E_\text{$\Gamma$-L}\sim 25$~meV. Moreover, in GaSb-based materials, the density of states and the electron effective mass are lower in the $\Gamma$ states than in the $L$ states~\cite{Snow1989,Vurgaftman2001}. As a consequence, a fraction of hot electrons can be scattered from $\Gamma$ to $L$ states at the QW, with $\Gamma \to L$ scattering times of the order of $10^2$~fs~\cite{Pelouch1995,Pellemans1996,Smith1998}. This short lifetime can reduce $\tau_\text{e}$ with respect to $\tau_0$.

Figure~\ref{Figure5} (a) also shows that, at intermediate temporal ranges (gray region), electron-hole pair relaxation processes prevail, inducing a fast decay characterized by $\left(\tau_{\text{e-h}}\right)^{-1} \equiv \left(\tau_\text{e}\right)^{-1} + \left(\tau_\text{h}\right)^{-1}$. Yet, at longer temporal ranges (dark gray region), the slow decay is dominated by the longer minority carrier lifetime $\tau_\text{h}$. It is worth noting that the tail obtained in these calculations at longer temporal ranges is a trace of residual minority holes that the approximation of $\tau_0 \to \infty$ precludes from being depleted. Consequently, the boundary of the time window considered in these simulations under such an approximation and particularly at resonance (highest transmission) must be considered with caution since during the experiments, residual holes can be mostly exhausted because of the recombination with resonance-transmitted electrons, making it difficult to observe this tail in the transient response.

Simulations also indicate that when the intensity of the calculated transient response diminishes due to low transmission probabilities, as represented by light blue lines, the fast electron-hole pair relaxation dynamics prevail even at shorter temporal ranges (light gray region). Conversely, high intensities provoked by high transmission probabilities as indicated by the black line, not only reinforce the filling process inside the double-barrier QW as depicted by the plateau in the light gray temporal region but also generate a larger amount of remnant carriers at long temporal ranges, which raises the tail of the transient response as observed in the dark gray region. The presence of these carriers can increase the electroluminescence background observed in the experiments (not shown here).

\begin{figure*}
	\includegraphics[width=17.8cm]{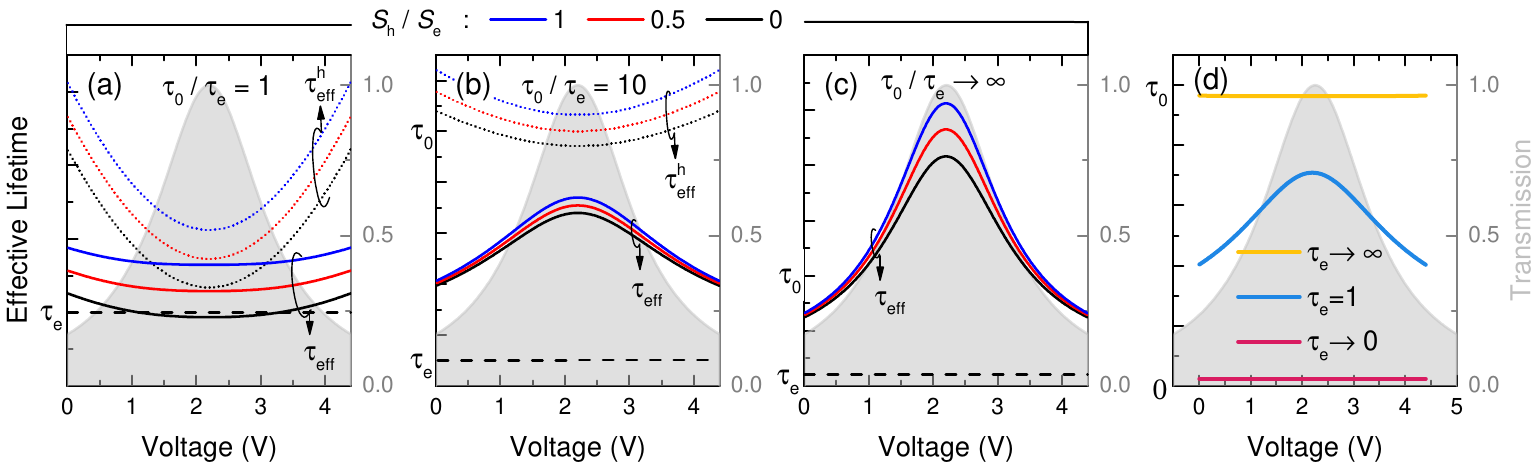}
	\caption{\label{Figure6} Calculated effective lifetime as a function of the applied voltage for (a) $\tau_0/\tau_\text{e}=1$, (b) $\tau_0/\tau_\text{e}=10$, and (c) $\tau_0/\tau_\text{e}\to\infty$. The influence of $S_\text{h}/S_\text{e}\to 0$ (black lines), $S_\text{h}/S_\text{e}=0.5$ (red lines), and $S_\text{h}/S_\text{e} = 1$ (blue lines) on the effective lifetimes is also presented. $\tau_\text{e}$ is indicated by a dashed black line. (d) Calculated effective lifetime as a function of the applied voltage for $\tau_\text{e}\to 0$ (red line), $\tau_\text{e}=1$ (blue line), and $\tau_\text{e}\to \infty$ (yellow line). A Lorentzian-like electronic transmission (gray) was assumed for all the calculations.}
\end{figure*}

In contrast, if $\tau_0 \sim \tau_\text{e}$ as presented in Fig.~\ref{Figure5} (b), only two time-scales can be reproduced, similarly to the transient response of RTD-As presented in Fig.~\ref{Figure5} (a). These time scales correspond to (i) a fast decay (light gray and gray regions) ruled by both electron-hole pair relaxation mechanisms and fast optical recombinations, which hampers filling processes inside the double-barrier QW, and (ii) a slow decay (dark gray regions) mostly influenced by hole lifetimes. In this case, thinner and lower barriers in addition to higher optical emission energies at the double-barrier QW reduce escaping times and radiative lifetimes. Consequently, the effective lifetime remains constant with the applied voltage, showing no modulation with the transmission.

\subsection{Effective lifetimes correlation with current conditions}
Since estimations of the experimental effective lifetimes are unavoidably performed by fitting procedures within finite temporal ranges, differences between these ranges in a transient response are crucial for the actual determination and characterization of various lifetimes. Under these conditions, the contribution of longer time scales that appears when the PL intensity is already very low, as occurs at the tails of the transient responses, can be neglected. However, it would be impossible to prevent the dependence of effective lifetimes on the transmission and consequently, on the current source for majority carriers $S_\text{e}$. This apparent drawback is actually an advantage of this procedure since three complementary dynamic conditions can be assessed under low and large currents.

If the intensity of the transient response is considered as an exponential decrease, the effective lifetime can be defined as a function of time by
\begin{equation}\label{tau_eff}
    \tau_\text{eff}(t) = -\left( \frac{d}{dt} \ln \mathscr{I}\right)^{-1},
\end{equation}
with $\mathscr{I}=np$ being the intensity of the emitted light. Thus, the evolution of $\tau_\text{eff}$ with transmission and time can be extracted from the calculated transient responses, as shown in the color-map of Fig.~\ref{Figure5} (c). Here, $\tau_0 / \tau_\text{e} = 10$ was assumed and line profiles of $\tau_\text{eff}$ for three different times, $t=2\tau_\text{e}$ (dashed line), $t=7\tau_\text{e}$ (dotted line), and $t=13\tau_\text{e}$ (dot-dashed line) allow to correlate results with the decay curves in panels (a) and (b). Data before $t=\tau_\text{e}$ (shaded region) could not be extracted due to the presence of the peak of maximum intensity in the transient responses. 

The results in Fig.~\ref{Figure5} (c) show the contrast between short, intermediate, and long temporal ranges, as well as the necessary conditions for accessing different time scales. In this way, slow dynamics characterized by high values of $\tau_\text{eff}$ (red) can be observed when the transmission increases at short ($t<5\tau_\text{e}$) and long ($t>10\tau_\text{e}$) temporal ranges, also represented by light and dark gray regions respectively in Figs.~\ref{Figure5} (a) and (b). However, at intermediate temporal ranges ($5\leq t/\tau_\text{e}\leq10$), $\tau_\text{eff}$ has no dependence on the transmission and reaches minimum values (purple) due to fast electron-hole pair relaxation mechanisms.

The dependence of the effective lifetime on transmission conditions enables the investigation of its correlation with the applied voltage. This correlation has been also emulated in Figs.~\ref{Figure6} (a)-(c) by solving the rate equation system for $\tau_0/\tau_\text{e}=1$, $\tau_0/\tau_\text{e}=10$, and $\tau_0/\tau_\text{e}\to\infty$, respectively. For the calculations, the transmission was assumed as a Lorentzian function of the voltage, as depicted by gray curves in the background, and the effective lifetimes were extracted from the calculated decay curves, at $t=2\tau_\text{e}$. The influence of varying the ratio between holes and electrons sources was also simulated for $S_\text{h}/S_\text{e}\to 0$ (black lines), $S_\text{h}/S_\text{e}=0.5$ (red lines), and $S_\text{h}/S_\text{e} = 1$ (blue lines).

Dotted lines correspond to the calculation of the effective lifetime just for holes $\tau^\text{h}_\text{eff}$, for $n$ set as constant in Eq.~\ref{tau_eff}. Calculations were carried out for the same $S_\text{h}/S_\text{e}$ ratios as before, as indicated by the colors of the dotted lines. The dependence of $\tau^\text{h}_\text{eff}$ with voltage allows assessing the holes' contribution to the carriers dynamics, independently on the role of electrons. This segmentation of the carriers contributions is necessary to evaluate their weight within the total effective lifetime.

Calculations show that, in all cases, the effective lifetime increases with $S_\text{h}/S_\text{e}$, once the minority relaxation dynamics gain relevance. In addition, a large current condition at resonance leads to two opposite effective lifetime responses under applied voltages. The first response is achieved when $\tau_0/\tau_\text{e}=1$ as depicted by Fig.~\ref{Figure6} (a). Here, effective lifetimes exhibit a dip at the resonance condition, similar to the dip presented by $\tau^\text{h}_\text{eff}$. Thus, holes relaxation dynamics govern the voltage dependence of $\tau_\text{eff}$, slowing down the effective lifetime to values closer to $\tau_\text{e}$, while out-of-resonance conditions reveal longer effective lifetimes, higher than $\tau_\text{e}$. In double-barrier QWs, where radiative lifetimes are comparable with the electron escape times, the filling-of-states process is inefficient, inducing carrier dynamics ruled by electron-hole relaxation processes during out-of-resonance conditions, and by the escape of electrons at resonance conditions.

The second type of response is originated when $\tau_0/\tau_\text{e}>1$. Large current conditions induce a peak in the effective lifetimes in this case, which become closer to $\tau_0$ for $\tau_0/\tau_\text{e}=10$, and to $\tau_\text{h}$ for $\tau_0/\tau_\text{e}\to\infty$, as indicated by Figs.~\ref{Figure6} (b) and (c), respectively. Low electric currents during out-of-resonance conditions reduce the effective lifetime close to $\tau_\text{e}$ since in this situation the dynamics are governed by the escape of majority electrons. However, high electric currents at resonance enhance electron-hole pair relaxation dynamics due to the filling-of-states or intervalley-scattering processes.

The influence of varying the electrons escape time on the voltage dependence of the effective lifetimes has been also explored. Figure~\ref{Figure6} (d) shows the results for $\tau_\text{e}\to 0$ (red line), $\tau_\text{e}=1$ (blue line) and $\tau_\text{e}\to\infty$ (yellow line). For the simulations, $S_\text{h}/S_\text{e}=0.5$, $\tau_\text{h}\sim 10^2$, and $\tau_0\sim 10$ have been considered. Simulations show that slow (fast) electrons escape processes, which means $\tau_\text{e} \to \infty$ ($\tau_\text{e} \to 0$) make optical recombination times (electrons lifetimes) prevail and thus, effective lifetimes present no dependence with the applied voltage. On the contrary, intermediate values of $\tau_\text{e}$ lead to a competition between optical recombination and electrons escaping processes which depend on the applied voltage, as depicted by the blue line. The former situations may correspond to RTD-As, where electrons relaxation dynamics are expected to be faster than in RTD-Sb, due to its lower barriers, corroborated by the observed short effective lifetimes.

\section{Conclusions}
We demonstrate how carrier dynamics in double-barrier QWs are ruled by three different time scales corresponding to radiative recombination processes, fast electron-hole pair relaxations, including intervalley scattering inside the double-barrier QW, and slow minority carrier relaxation mechanisms. Optical recombination time scales dominate in Sb-based systems under high coherent currents (resonance condition), as long as the electrons' escape time is shorter than optical recombination times. This is possible thanks to high band offsets and the small $\Gamma$-L energy separation present in Sb-based DBSs. These factors induce a longer electrons' escape time than in As-based DBSs, despite the similarities in the current densities of both systems. Fast dynamics are enhanced for low coherent currents (out-of-resonance conditions) while slow minority carrier relaxation can prevail at resonance conditions, when radiative recombination processes are too slow ($\tau_0 \to \infty$), thus increasing the observed effective lifetimes in Sb-based systems. The above-presented results account for the correlation between lifetimes and current-density-voltage characteristics, particularly at resonance conditions in Sb-based DBSs. Moreover, the lack of correlation between experimental effective lifetimes and the current density for out-of-resonance voltages indicates that the source of electrons, which affects the electron-hole pair dynamics, comes essentially from the electronic coherent channel. Consequently, tuning the current through the RTD allows assessing these time scales almost independently.

\begin{acknowledgments}
The authors are grateful for financial support by BAYLAT and by Brazilian agencies: the Fundação de Amparo à Pesquisa do Estado de São Paulo (FAPESP) - grants No. 2013/18719-1, No. 2014/07375-2, No. 2014/19142-2, No. 2014/02112-3, No. 2015/13771-0 and No. 2018/01914-0, the Conselho Nacional de Desenvolvimento Científico e Tecnológico (CNPq), the Coordenação de Aperfeiçonamento de Pessoal de Nível Superior -Brasil (CAPES)- Finance Code 001, and the MSCA-ITN-2020 QUANTIMONY from the European Union’s Horizon 2020 programme under Grant agreement ID: 956548.
\end{acknowledgments}

\bibliography{References}

\begin{thebibliography}{52}%
\makeatletter
\providecommand \@ifxundefined [1]{%
 \@ifx{#1\undefined}
}%
\providecommand \@ifnum [1]{%
 \ifnum #1\expandafter \@firstoftwo
 \else \expandafter \@secondoftwo
 \fi
}%
\providecommand \@ifx [1]{%
 \ifx #1\expandafter \@firstoftwo
 \else \expandafter \@secondoftwo
 \fi
}%
\providecommand \natexlab [1]{#1}%
\providecommand \enquote  [1]{``#1''}%
\providecommand \bibnamefont  [1]{#1}%
\providecommand \bibfnamefont [1]{#1}%
\providecommand \citenamefont [1]{#1}%
\providecommand \href@noop [0]{\@secondoftwo}%
\providecommand \href [0]{\begingroup \@sanitize@url \@href}%
\providecommand \@href[1]{\@@startlink{#1}\@@href}%
\providecommand \@@href[1]{\endgroup#1\@@endlink}%
\providecommand \@sanitize@url [0]{\catcode `\\12\catcode `\$12\catcode
  `\&12\catcode `\#12\catcode `\^12\catcode `\_12\catcode `\%12\relax}%
\providecommand \@@startlink[1]{}%
\providecommand \@@endlink[0]{}%
\providecommand \url  [0]{\begingroup\@sanitize@url \@url }%
\providecommand \@url [1]{\endgroup\@href {#1}{\urlprefix }}%
\providecommand \urlprefix  [0]{URL }%
\providecommand \Eprint [0]{\href }%
\providecommand \doibase [0]{https://doi.org/}%
\providecommand \selectlanguage [0]{\@gobble}%
\providecommand \bibinfo  [0]{\@secondoftwo}%
\providecommand \bibfield  [0]{\@secondoftwo}%
\providecommand \translation [1]{[#1]}%
\providecommand \BibitemOpen [0]{}%
\providecommand \bibitemStop [0]{}%
\providecommand \bibitemNoStop [0]{.\EOS\space}%
\providecommand \EOS [0]{\spacefactor3000\relax}%
\providecommand \BibitemShut  [1]{\csname bibitem#1\endcsname}%
\let\auto@bib@innerbib\@empty
%</preamble>
\bibitem [{\citenamefont {Soderstrom}\ \emph {et~al.}(1990)\citenamefont
  {Soderstrom}, \citenamefont {Chow},\ and\ \citenamefont
  {McGill}}]{Soderstrom1990}%
  \BibitemOpen
  \bibfield  {author} {\bibinfo {author} {\bibfnamefont {J.}~\bibnamefont
  {Soderstrom}}, \bibinfo {author} {\bibfnamefont {D.}~\bibnamefont {Chow}},\
  and\ \bibinfo {author} {\bibfnamefont {T.}~\bibnamefont {McGill}},\
  }\bibfield  {title} {\bibinfo {title} {{InAs/AlSb double-barrier structure
  with large peak-to-valley current ratio: a candidate for high-frequency
  microwave devices}},\ }\href {https://doi.org/10.1109/55.46920} {\bibfield
  {journal} {\bibinfo  {journal} {IEEE Electron Device Lett.}\ }\textbf
  {\bibinfo {volume} {11}},\ \bibinfo {pages} {27} (\bibinfo {year}
  {1990})}\BibitemShut {NoStop}%
\bibitem [{\citenamefont {Feiginov}\ \emph {et~al.}(2014)\citenamefont
  {Feiginov}, \citenamefont {Kanaya}, \citenamefont {Suzuki},\ and\
  \citenamefont {Asada}}]{Feiginov}%
  \BibitemOpen
  \bibfield  {author} {\bibinfo {author} {\bibfnamefont {M.}~\bibnamefont
  {Feiginov}}, \bibinfo {author} {\bibfnamefont {H.}~\bibnamefont {Kanaya}},
  \bibinfo {author} {\bibfnamefont {S.}~\bibnamefont {Suzuki}},\ and\ \bibinfo
  {author} {\bibfnamefont {M.}~\bibnamefont {Asada}},\ }\bibfield  {title}
  {\bibinfo {title} {{Operation of resonant-tunneling diodes with strong back
  injection from the collector at frequencies up to 1.46 THz}},\ }\href
  {https://doi.org/10.1063/1.4884602} {\bibfield  {journal} {\bibinfo
  {journal} {Appl. Phys. Lett.}\ }\textbf {\bibinfo {volume} {104}},\ \bibinfo
  {pages} {243509} (\bibinfo {year} {2014})}\BibitemShut {NoStop}%
\bibitem [{\citenamefont {Muttlak}\ \emph {et~al.}(2018)\citenamefont
  {Muttlak}, \citenamefont {Abdulwahid}, \citenamefont {Sexton}, \citenamefont
  {Kelly},\ and\ \citenamefont {Missous}}]{Muttlak}%
  \BibitemOpen
  \bibfield  {author} {\bibinfo {author} {\bibfnamefont {S.~G.}\ \bibnamefont
  {Muttlak}}, \bibinfo {author} {\bibfnamefont {O.~S.}\ \bibnamefont
  {Abdulwahid}}, \bibinfo {author} {\bibfnamefont {J.}~\bibnamefont {Sexton}},
  \bibinfo {author} {\bibfnamefont {M.~J.}\ \bibnamefont {Kelly}},\ and\
  \bibinfo {author} {\bibfnamefont {M.}~\bibnamefont {Missous}},\ }\bibfield
  {title} {\bibinfo {title} {{InGaAs/AlAs Resonant Tunneling Diodes for THz
  Applications: An Experimental Investigation}},\ }\href
  {https://doi.org/10.1109/JEDS.2018.2797951} {\bibfield  {journal} {\bibinfo
  {journal} {IEEE J. Electron Devices Soc.}\ }\textbf {\bibinfo {volume} {6}},\
  \bibinfo {pages} {254} (\bibinfo {year} {2018})}\BibitemShut {NoStop}%
\bibitem [{\citenamefont {Tavares}\ \emph {et~al.}(2017)\citenamefont
  {Tavares}, \citenamefont {Pessoa}, \citenamefont {Figueiredo},\ and\
  \citenamefont {Salgado}}]{Tavares2017}%
  \BibitemOpen
  \bibfield  {author} {\bibinfo {author} {\bibfnamefont {J.~S.}\ \bibnamefont
  {Tavares}}, \bibinfo {author} {\bibfnamefont {L.~M.}\ \bibnamefont {Pessoa}},
  \bibinfo {author} {\bibfnamefont {J.~M.~L.}\ \bibnamefont {Figueiredo}},\
  and\ \bibinfo {author} {\bibfnamefont {H.~M.}\ \bibnamefont {Salgado}},\
  }\bibfield  {title} {\bibinfo {title} {{Analysis of resonant tunnelling diode
  oscillators under optical modulation}},\ }in\ \href
  {https://doi.org/10.1109/ICTON.2017.8024755} {\emph {\bibinfo {booktitle}
  {2017 19th International Conference on Transparent Optical Networks
  (ICTON)}}}\ (\bibinfo {year} {2017})\ pp.\ \bibinfo {pages}
  {1--4}\BibitemShut {NoStop}%
\bibitem [{\citenamefont {Zhang}\ \emph {et~al.}(2018)\citenamefont {Zhang},
  \citenamefont {Watson}, \citenamefont {Wang}, \citenamefont {Figueiredo},
  \citenamefont {Wasige},\ and\ \citenamefont {Kelly}}]{Zhang2018}%
  \BibitemOpen
  \bibfield  {author} {\bibinfo {author} {\bibfnamefont {W.}~\bibnamefont
  {Zhang}}, \bibinfo {author} {\bibfnamefont {S.}~\bibnamefont {Watson}},
  \bibinfo {author} {\bibfnamefont {J.}~\bibnamefont {Wang}}, \bibinfo {author}
  {\bibfnamefont {J.}~\bibnamefont {Figueiredo}}, \bibinfo {author}
  {\bibfnamefont {E.}~\bibnamefont {Wasige}},\ and\ \bibinfo {author}
  {\bibfnamefont {A.~E.}\ \bibnamefont {Kelly}},\ }\bibfield  {title} {\bibinfo
  {title} {{Optical Characteristics Analysis of Resonant Tunneling Diode
  Photodiode Based Oscillators}},\ }in\ \href
  {https://doi.org/10.1109/VTCSpring.2018.8417889} {\emph {\bibinfo {booktitle}
  {2018 IEEE 87th Vehicular Technology Conference (VTC Spring)}}}\ (\bibinfo
  {year} {2018})\ pp.\ \bibinfo {pages} {1--6}\BibitemShut {NoStop}%
\bibitem [{\citenamefont {Watson}\ \emph {et~al.}(2019)\citenamefont {Watson},
  \citenamefont {Zhang}, \citenamefont {Tavares}, \citenamefont {Figueiredo},
  \citenamefont {Cantu}, \citenamefont {Wang}, \citenamefont {Wasige},
  \citenamefont {Salgado}, \citenamefont {Pessoa},\ and\ \citenamefont
  {Kelly}}]{Watson2019}%
  \BibitemOpen
  \bibfield  {author} {\bibinfo {author} {\bibfnamefont {S.}~\bibnamefont
  {Watson}}, \bibinfo {author} {\bibfnamefont {W.}~\bibnamefont {Zhang}},
  \bibinfo {author} {\bibfnamefont {J.}~\bibnamefont {Tavares}}, \bibinfo
  {author} {\bibfnamefont {J.}~\bibnamefont {Figueiredo}}, \bibinfo {author}
  {\bibfnamefont {H.}~\bibnamefont {Cantu}}, \bibinfo {author} {\bibfnamefont
  {J.}~\bibnamefont {Wang}}, \bibinfo {author} {\bibfnamefont {E.}~\bibnamefont
  {Wasige}}, \bibinfo {author} {\bibfnamefont {H.}~\bibnamefont {Salgado}},
  \bibinfo {author} {\bibfnamefont {L.}~\bibnamefont {Pessoa}},\ and\ \bibinfo
  {author} {\bibfnamefont {A.}~\bibnamefont {Kelly}},\ }\bibfield  {title}
  {\bibinfo {title} {{Resonant tunneling diode photodetectors for optical
  communications}},\ }\href {https://doi.org/https://doi.org/10.1002/mop.31689}
  {\bibfield  {journal} {\bibinfo  {journal} {Microw. Opt. Technol. Lett.}\
  }\textbf {\bibinfo {volume} {61}},\ \bibinfo {pages} {1121} (\bibinfo {year}
  {2019})}\BibitemShut {NoStop}%
\bibitem [{\citenamefont {Wang}\ \emph {et~al.}(2022)\citenamefont {Wang},
  \citenamefont {Naftaly},\ and\ \citenamefont {Wasige}}]{Wang2022}%
  \BibitemOpen
  \bibfield  {author} {\bibinfo {author} {\bibfnamefont {J.}~\bibnamefont
  {Wang}}, \bibinfo {author} {\bibfnamefont {M.}~\bibnamefont {Naftaly}},\ and\
  \bibinfo {author} {\bibfnamefont {E.}~\bibnamefont {Wasige}},\ }\bibfield
  {title} {\bibinfo {title} {{An Overview of Terahertz Imaging with Resonant
  Tunneling Diodes}},\ }\href {https://www.mdpi.com/2076-3417/12/8/3822}
  {\bibfield  {journal} {\bibinfo  {journal} {Appl. Sci.}\ }\textbf {\bibinfo
  {volume} {12}} (\bibinfo {year} {2022})}\BibitemShut {NoStop}%
\bibitem [{\citenamefont {Buttiker}(1988)}]{Buttiker1988}%
  \BibitemOpen
  \bibfield  {author} {\bibinfo {author} {\bibfnamefont {M.}~\bibnamefont
  {Buttiker}},\ }\bibfield  {title} {\bibinfo {title} {{Coherent and sequential
  tunneling in series barriers}},\ }\href {https://doi.org/10.1147/rd.321.0063}
  {\bibfield  {journal} {\bibinfo  {journal} {IBM J. Res. Dev.}\ }\textbf
  {\bibinfo {volume} {32}},\ \bibinfo {pages} {63} (\bibinfo {year}
  {1988})}\BibitemShut {NoStop}%
\bibitem [{\citenamefont {Jimenez}\ \emph {et~al.}(1994)\citenamefont
  {Jimenez}, \citenamefont {Li},\ and\ \citenamefont {Wang}}]{Jimenez1994}%
  \BibitemOpen
  \bibfield  {author} {\bibinfo {author} {\bibfnamefont {J.~L.}\ \bibnamefont
  {Jimenez}}, \bibinfo {author} {\bibfnamefont {X.}~\bibnamefont {Li}},\ and\
  \bibinfo {author} {\bibfnamefont {W.~I.}\ \bibnamefont {Wang}},\ }\bibfield
  {title} {\bibinfo {title} {{Resonant tunneling in AlSb‐GaSb‐AlSb and
  AlSb‐InGaSb‐AlSb double barrier heterostructures}},\ }\href
  {https://doi.org/10.1063/1.111705} {\bibfield  {journal} {\bibinfo  {journal}
  {Appl. Phys. Lett.}\ }\textbf {\bibinfo {volume} {64}},\ \bibinfo {pages}
  {2127} (\bibinfo {year} {1994})}\BibitemShut {NoStop}%
\bibitem [{\citenamefont {Jimenez}\ \emph {et~al.}(1995)\citenamefont
  {Jimenez}, \citenamefont {Mendez}, \citenamefont {Li},\ and\ \citenamefont
  {Wang}}]{Jimenez1995}%
  \BibitemOpen
  \bibfield  {author} {\bibinfo {author} {\bibfnamefont {J.~L.}\ \bibnamefont
  {Jimenez}}, \bibinfo {author} {\bibfnamefont {E.~E.}\ \bibnamefont {Mendez}},
  \bibinfo {author} {\bibfnamefont {X.}~\bibnamefont {Li}},\ and\ \bibinfo
  {author} {\bibfnamefont {W.~I.}\ \bibnamefont {Wang}},\ }\bibfield  {title}
  {\bibinfo {title} {{Intrinsic bistability by charge accumulation in an
  L-valley state in GaSb-AlSb resonant-tunneling diodes}},\ }\href
  {https://doi.org/10.1103/PhysRevB.52.R5495} {\bibfield  {journal} {\bibinfo
  {journal} {Phys. Rev. B}\ }\textbf {\bibinfo {volume} {52}},\ \bibinfo
  {pages} {R5495} (\bibinfo {year} {1995})}\BibitemShut {NoStop}%
\bibitem [{\citenamefont {Lee}\ \emph {et~al.}(2018)\citenamefont {Lee},
  \citenamefont {Shin}, \citenamefont {Byun},\ and\ \citenamefont
  {Kim}}]{Lee2018}%
  \BibitemOpen
  \bibfield  {author} {\bibinfo {author} {\bibfnamefont {J.-H.}\ \bibnamefont
  {Lee}}, \bibinfo {author} {\bibfnamefont {M.}~\bibnamefont {Shin}}, \bibinfo
  {author} {\bibfnamefont {S.-J.}\ \bibnamefont {Byun}},\ and\ \bibinfo
  {author} {\bibfnamefont {W.}~\bibnamefont {Kim}},\ }\bibfield  {title}
  {\bibinfo {title} {{Wigner Transport Simulation of Resonant Tunneling Diodes
  with Auxiliary Quantum Wells}},\ }\href {https://doi.org/10.3938/jkps.72.622}
  {\bibfield  {journal} {\bibinfo  {journal} {J. Korean Phys. Soc.}\ }\textbf
  {\bibinfo {volume} {72}},\ \bibinfo {pages} {622} (\bibinfo {year}
  {2018})}\BibitemShut {NoStop}%
\bibitem [{\citenamefont {Brown}\ \emph {et~al.}(2021)\citenamefont {Brown},
  \citenamefont {Zhang}, \citenamefont {Growden}, \citenamefont {Fakhimi},\
  and\ \citenamefont {Berger}}]{Brown2021}%
  \BibitemOpen
  \bibfield  {author} {\bibinfo {author} {\bibfnamefont {E.}~\bibnamefont
  {Brown}}, \bibinfo {author} {\bibfnamefont {W.-D.}\ \bibnamefont {Zhang}},
  \bibinfo {author} {\bibfnamefont {T.}~\bibnamefont {Growden}}, \bibinfo
  {author} {\bibfnamefont {P.}~\bibnamefont {Fakhimi}},\ and\ \bibinfo {author}
  {\bibfnamefont {P.}~\bibnamefont {Berger}},\ }\bibfield  {title} {\bibinfo
  {title} {{Electroluminescence in Unipolar-Doped
  ${\mathrm{In}}_{0.53}{\mathrm{Ga}}_{0.47}\mathrm{As}/\mathrm{Al}\mathrm{As}$
  Resonant-Tunneling Diodes: A Competition between Interband Tunneling and
  Impact Ionization}},\ }\href
  {https://doi.org/10.1103/PhysRevApplied.16.054008} {\bibfield  {journal}
  {\bibinfo  {journal} {Phys. Rev. Applied}\ }\textbf {\bibinfo {volume}
  {16}},\ \bibinfo {pages} {054008} (\bibinfo {year} {2021})}\BibitemShut
  {NoStop}%
\bibitem [{\citenamefont {Sun}\ \emph {et~al.}(1998)\citenamefont {Sun},
  \citenamefont {Haddad}, \citenamefont {Mazumder},\ and\ \citenamefont
  {Schulman}}]{Sun}%
  \BibitemOpen
  \bibfield  {author} {\bibinfo {author} {\bibfnamefont {J.~P.}\ \bibnamefont
  {Sun}}, \bibinfo {author} {\bibfnamefont {G.}~\bibnamefont {Haddad}},
  \bibinfo {author} {\bibfnamefont {P.}~\bibnamefont {Mazumder}},\ and\
  \bibinfo {author} {\bibfnamefont {J.}~\bibnamefont {Schulman}},\ }\bibfield
  {title} {\bibinfo {title} {{Resonant tunneling diodes: models and
  properties}},\ }\href {https://doi.org/10.1109/5.663541} {\bibfield
  {journal} {\bibinfo  {journal} {Proc. IEEE}\ }\textbf {\bibinfo {volume}
  {86}},\ \bibinfo {pages} {641} (\bibinfo {year} {1998})}\BibitemShut
  {NoStop}%
\bibitem [{\citenamefont {Muttlak}\ \emph {et~al.}(2019)\citenamefont
  {Muttlak}, \citenamefont {Abdulwahid}, \citenamefont {Sexton},\ and\
  \citenamefont {Missous}}]{Muttlak2019}%
  \BibitemOpen
  \bibfield  {author} {\bibinfo {author} {\bibfnamefont {S.~G.}\ \bibnamefont
  {Muttlak}}, \bibinfo {author} {\bibfnamefont {O.}~\bibnamefont {Abdulwahid}},
  \bibinfo {author} {\bibfnamefont {J.}~\bibnamefont {Sexton}},\ and\ \bibinfo
  {author} {\bibfnamefont {M.}~\bibnamefont {Missous}},\ }\bibfield  {title}
  {\bibinfo {title} {{InGaAs/AlAs Resonant Tunneling Diodes with Very High
  Negative Differential Conductance for Efficient and Cost-Effective
  mm-Wave/THz Sources}},\ }in\ \href
  {https://doi.org/10.1109/UCMMT47867.2019.9008338} {\emph {\bibinfo
  {booktitle} {2019 12th UK-Europe-China Workshop on Millimeter Waves and
  Terahertz Technologies (UCMMT)}}}\ (\bibinfo {year} {2019})\ pp.\ \bibinfo
  {pages} {1--3}\BibitemShut {NoStop}%
\bibitem [{\citenamefont {Guarin~Castro}\ \emph {et~al.}(2020)\citenamefont
  {Guarin~Castro}, \citenamefont {Rothmayr}, \citenamefont {Krüger},
  \citenamefont {Knebl}, \citenamefont {Schade}, \citenamefont {Koeth},
  \citenamefont {Worschech}, \citenamefont {Lopez-Richard}, \citenamefont
  {Marques}, \citenamefont {Hartmann}, \citenamefont {Pfenning},\ and\
  \citenamefont {Höfling}}]{Guarin2020}%
  \BibitemOpen
  \bibfield  {author} {\bibinfo {author} {\bibfnamefont {E.~D.}\ \bibnamefont
  {Guarin~Castro}}, \bibinfo {author} {\bibfnamefont {F.}~\bibnamefont
  {Rothmayr}}, \bibinfo {author} {\bibfnamefont {S.}~\bibnamefont {Krüger}},
  \bibinfo {author} {\bibfnamefont {G.}~\bibnamefont {Knebl}}, \bibinfo
  {author} {\bibfnamefont {A.}~\bibnamefont {Schade}}, \bibinfo {author}
  {\bibfnamefont {J.}~\bibnamefont {Koeth}}, \bibinfo {author} {\bibfnamefont
  {L.}~\bibnamefont {Worschech}}, \bibinfo {author} {\bibfnamefont
  {V.}~\bibnamefont {Lopez-Richard}}, \bibinfo {author} {\bibfnamefont {G.~E.}\
  \bibnamefont {Marques}}, \bibinfo {author} {\bibfnamefont {F.}~\bibnamefont
  {Hartmann}}, \bibinfo {author} {\bibfnamefont {A.}~\bibnamefont {Pfenning}},\
  and\ \bibinfo {author} {\bibfnamefont {S.}~\bibnamefont {Höfling}},\
  }\bibfield  {title} {\bibinfo {title} {{Resonant tunneling of electrons in
  AlSb/GaInAsSb double barrier quantum wells}},\ }\href
  {https://doi.org/10.1063/5.0008959} {\bibfield  {journal} {\bibinfo
  {journal} {AIP Adv.}\ }\textbf {\bibinfo {volume} {10}},\ \bibinfo {pages}
  {055024} (\bibinfo {year} {2020})}\BibitemShut {NoStop}%
\bibitem [{\citenamefont {Rebey}\ \emph {et~al.}(2022)\citenamefont {Rebey},
  \citenamefont {Mbarki}, \citenamefont {Rebei},\ and\ \citenamefont
  {Messaoudi}}]{Rebey2022}%
  \BibitemOpen
  \bibfield  {author} {\bibinfo {author} {\bibfnamefont {A.}~\bibnamefont
  {Rebey}}, \bibinfo {author} {\bibfnamefont {M.}~\bibnamefont {Mbarki}},
  \bibinfo {author} {\bibfnamefont {H.}~\bibnamefont {Rebei}},\ and\ \bibinfo
  {author} {\bibfnamefont {S.}~\bibnamefont {Messaoudi}},\ }\bibfield  {title}
  {\bibinfo {title} {{Performance optimization of AlGaAs/GaAsBiN resonant
  tunneling diode}},\ }\href
  {https://doi.org/https://doi.org/10.1016/j.ijleo.2022.169793} {\bibfield
  {journal} {\bibinfo  {journal} {Optik}\ }\textbf {\bibinfo {volume} {268}},\
  \bibinfo {pages} {169793} (\bibinfo {year} {2022})}\BibitemShut {NoStop}%
\bibitem [{\citenamefont {Takeda}\ \emph {et~al.}(2015)\citenamefont {Takeda},
  \citenamefont {Ichiki}, \citenamefont {Kusano}, \citenamefont {Sugimoto},\
  and\ \citenamefont {Motohiro}}]{Takeda2015}%
  \BibitemOpen
  \bibfield  {author} {\bibinfo {author} {\bibfnamefont {Y.}~\bibnamefont
  {Takeda}}, \bibinfo {author} {\bibfnamefont {A.}~\bibnamefont {Ichiki}},
  \bibinfo {author} {\bibfnamefont {Y.}~\bibnamefont {Kusano}}, \bibinfo
  {author} {\bibfnamefont {N.}~\bibnamefont {Sugimoto}},\ and\ \bibinfo
  {author} {\bibfnamefont {T.}~\bibnamefont {Motohiro}},\ }\bibfield  {title}
  {\bibinfo {title} {{Resonant tunneling diodes as energy-selective contacts
  used in hot-carrier solar cells}},\ }\href
  {https://doi.org/10.1063/1.4931888} {\bibfield  {journal} {\bibinfo
  {journal} {J. Appl. Phys.}\ }\textbf {\bibinfo {volume} {118}},\ \bibinfo
  {pages} {124510} (\bibinfo {year} {2015})}\BibitemShut {NoStop}%
\bibitem [{\citenamefont {Jehl}\ \emph {et~al.}(2017)\citenamefont {Jehl},
  \citenamefont {Suchet}, \citenamefont {Julian}, \citenamefont {Bernard},
  \citenamefont {Miyashita}, \citenamefont {Gibelli}, \citenamefont {Okada},\
  and\ \citenamefont {Guillemolles}}]{Zacharie2017}%
  \BibitemOpen
  \bibfield  {author} {\bibinfo {author} {\bibfnamefont {Z.}~\bibnamefont
  {Jehl}}, \bibinfo {author} {\bibfnamefont {D.}~\bibnamefont {Suchet}},
  \bibinfo {author} {\bibfnamefont {A.}~\bibnamefont {Julian}}, \bibinfo
  {author} {\bibfnamefont {C.}~\bibnamefont {Bernard}}, \bibinfo {author}
  {\bibfnamefont {N.}~\bibnamefont {Miyashita}}, \bibinfo {author}
  {\bibfnamefont {F.}~\bibnamefont {Gibelli}}, \bibinfo {author} {\bibfnamefont
  {Y.}~\bibnamefont {Okada}},\ and\ \bibinfo {author} {\bibfnamefont {J.-F.}\
  \bibnamefont {Guillemolles}},\ }\bibfield  {title} {\bibinfo {title}
  {{Modeling and characterization of double resonant tunneling diodes for
  application as energy selective contacts in hot carrier solar cells}},\ }in\
  \href {https://doi.org/10.1117/12.2250473} {\emph {\bibinfo {booktitle}
  {Physics, Simulation, and Photonic Engineering of Photovoltaic Devices
  VI}}},\ Vol.\ \bibinfo {volume} {10099},\ \bibinfo {editor} {edited by\
  \bibinfo {editor} {\bibfnamefont {A.}~\bibnamefont {Freundlich}}, \bibinfo
  {editor} {\bibfnamefont {L.}~\bibnamefont {Lombez}},\ and\ \bibinfo {editor}
  {\bibfnamefont {M.}~\bibnamefont {Sugiyama}}},\ \bibinfo {organization}
  {International Society for Optics and Photonics}\ (\bibinfo  {publisher}
  {SPIE},\ \bibinfo {year} {2017})\ p.\ \bibinfo {pages} {100990N}\BibitemShut
  {NoStop}%
\bibitem [{\citenamefont {Suchet}\ \emph {et~al.}(2017)\citenamefont {Suchet},
  \citenamefont {Jehl}, \citenamefont {Okada},\ and\ \citenamefont
  {Guillemoles}}]{Suchet2017}%
  \BibitemOpen
  \bibfield  {author} {\bibinfo {author} {\bibfnamefont {D.}~\bibnamefont
  {Suchet}}, \bibinfo {author} {\bibfnamefont {Z.}~\bibnamefont {Jehl}},
  \bibinfo {author} {\bibfnamefont {Y.}~\bibnamefont {Okada}},\ and\ \bibinfo
  {author} {\bibfnamefont {J.-F.}\ \bibnamefont {Guillemoles}},\ }\bibfield
  {title} {\bibinfo {title} {{Influence of Hot-Carrier Extraction from a
  Photovoltaic Absorber: An Evaporative Approach}},\ }\href
  {https://doi.org/10.1103/PhysRevApplied.8.034030} {\bibfield  {journal}
  {\bibinfo  {journal} {Phys. Rev. Applied}\ }\textbf {\bibinfo {volume} {8}},\
  \bibinfo {pages} {034030} (\bibinfo {year} {2017})}\BibitemShut {NoStop}%
\bibitem [{\citenamefont {Harada}\ and\ \citenamefont
  {Kuroda}(1986)}]{Harada1986}%
  \BibitemOpen
  \bibfield  {author} {\bibinfo {author} {\bibfnamefont {N.}~\bibnamefont
  {Harada}}\ and\ \bibinfo {author} {\bibfnamefont {S.}~\bibnamefont
  {Kuroda}},\ }\bibfield  {title} {\bibinfo {title} {{Lifetime of Resonant
  State in a Resonant Tunneling System}},\ }\href
  {https://doi.org/10.1143/jjap.25.l871} {\bibfield  {journal} {\bibinfo
  {journal} {Jpn. J. Appl. Phys.}\ }\textbf {\bibinfo {volume} {25}},\ \bibinfo
  {pages} {L871} (\bibinfo {year} {1986})}\BibitemShut {NoStop}%
\bibitem [{\citenamefont {Tsuchiya}\ \emph {et~al.}(1987)\citenamefont
  {Tsuchiya}, \citenamefont {Matsusue},\ and\ \citenamefont
  {Sakaki}}]{Tsuchiya1987}%
  \BibitemOpen
  \bibfield  {author} {\bibinfo {author} {\bibfnamefont {M.}~\bibnamefont
  {Tsuchiya}}, \bibinfo {author} {\bibfnamefont {T.}~\bibnamefont {Matsusue}},\
  and\ \bibinfo {author} {\bibfnamefont {H.}~\bibnamefont {Sakaki}},\
  }\bibfield  {title} {\bibinfo {title} {{Tunneling escape rate of electrons
  from quantum well in double-barrier heterostructures}},\ }\href
  {https://doi.org/10.1103/PhysRevLett.59.2356} {\bibfield  {journal} {\bibinfo
   {journal} {Phys. Rev. Lett.}\ }\textbf {\bibinfo {volume} {59}},\ \bibinfo
  {pages} {2356} (\bibinfo {year} {1987})}\BibitemShut {NoStop}%
\bibitem [{\citenamefont {Käß}\ \emph {et~al.}(1998)\citenamefont {Käß},
  \citenamefont {Schuddinck}, \citenamefont {Goovaerts}, \citenamefont {{Van
  Hoof}},\ and\ \citenamefont {Borghs}}]{Kas1998}%
  \BibitemOpen
  \bibfield  {author} {\bibinfo {author} {\bibfnamefont {H.}~\bibnamefont
  {Käß}}, \bibinfo {author} {\bibfnamefont {W.}~\bibnamefont {Schuddinck}},
  \bibinfo {author} {\bibfnamefont {E.}~\bibnamefont {Goovaerts}}, \bibinfo
  {author} {\bibfnamefont {C.}~\bibnamefont {{Van Hoof}}},\ and\ \bibinfo
  {author} {\bibfnamefont {G.}~\bibnamefont {Borghs}},\ }\bibfield  {title}
  {\bibinfo {title} {{Photoluminescence investigation of the tunnelling
  dynamics of holes and electrons in a p-type AlAs/GaAs resonant tunnelling
  structure}},\ }\href
  {https://doi.org/https://doi.org/10.1016/S0167-9317(98)00185-3} {\bibfield
  {journal} {\bibinfo  {journal} {Microelectron. Eng.}\ }\textbf {\bibinfo
  {volume} {43-44}},\ \bibinfo {pages} {355} (\bibinfo {year}
  {1998})}\BibitemShut {NoStop}%
\bibitem [{\citenamefont {Teran}\ \emph {et~al.}(2009)\citenamefont {Teran},
  \citenamefont {Mart{\'{\i}}n}, \citenamefont {Calleja}, \citenamefont
  {Vi{\~{n}}a}, \citenamefont {Eaves},\ and\ \citenamefont
  {Henini}}]{Teran_2009}%
  \BibitemOpen
  \bibfield  {author} {\bibinfo {author} {\bibfnamefont {F.~J.}\ \bibnamefont
  {Teran}}, \bibinfo {author} {\bibfnamefont {M.~D.}\ \bibnamefont
  {Mart{\'{\i}}n}}, \bibinfo {author} {\bibfnamefont {J.~M.}\ \bibnamefont
  {Calleja}}, \bibinfo {author} {\bibfnamefont {L.}~\bibnamefont {Vi{\~{n}}a}},
  \bibinfo {author} {\bibfnamefont {L.}~\bibnamefont {Eaves}},\ and\ \bibinfo
  {author} {\bibfnamefont {M.}~\bibnamefont {Henini}},\ }\bibfield  {title}
  {\bibinfo {title} {{Carrier injection effects on exciton dynamics in
  {GaAs}/{AlAs} resonant-tunneling diodes}},\ }\href
  {https://doi.org/10.1209/0295-5075/85/67010} {\bibfield  {journal} {\bibinfo
  {journal} {Europhys. Lett.}\ }\textbf {\bibinfo {volume} {85}},\ \bibinfo
  {pages} {67010} (\bibinfo {year} {2009})}\BibitemShut {NoStop}%
\bibitem [{\citenamefont {{Birner}}\ \emph {et~al.}(2007)\citenamefont
  {{Birner}}, \citenamefont {{Zibold}}, \citenamefont {{Andlauer}},
  \citenamefont {{Kubis}}, \citenamefont {{Sabathil}}, \citenamefont
  {{Trellakis}},\ and\ \citenamefont {{Vogl}}}]{Birner2007}%
  \BibitemOpen
  \bibfield  {author} {\bibinfo {author} {\bibfnamefont {S.}~\bibnamefont
  {{Birner}}}, \bibinfo {author} {\bibfnamefont {T.}~\bibnamefont {{Zibold}}},
  \bibinfo {author} {\bibfnamefont {T.}~\bibnamefont {{Andlauer}}}, \bibinfo
  {author} {\bibfnamefont {T.}~\bibnamefont {{Kubis}}}, \bibinfo {author}
  {\bibfnamefont {M.}~\bibnamefont {{Sabathil}}}, \bibinfo {author}
  {\bibfnamefont {A.}~\bibnamefont {{Trellakis}}},\ and\ \bibinfo {author}
  {\bibfnamefont {P.}~\bibnamefont {{Vogl}}},\ }\bibfield  {title} {\bibinfo
  {title} {{nextnano: General purpose 3-D simulations}},\ }\href
  {https://doi.org/10.1109/TED.2007.902871} {\bibfield  {journal} {\bibinfo
  {journal} {IEEE Trans. Electron Devices}\ }\textbf {\bibinfo {volume} {54}},\
  \bibinfo {pages} {2137} (\bibinfo {year} {2007})}\BibitemShut {NoStop}%
\bibitem [{\citenamefont {Cardozo~de Oliveira}\ \emph
  {et~al.}(2021)\citenamefont {Cardozo~de Oliveira}, \citenamefont {Naranjo},
  \citenamefont {Pfenning}, \citenamefont {Lopez-Richard}, \citenamefont
  {Marques}, \citenamefont {Worschech}, \citenamefont {Hartmann}, \citenamefont
  {H\"ofling},\ and\ \citenamefont {Teodoro}}]{Cardozo2021}%
  \BibitemOpen
  \bibfield  {author} {\bibinfo {author} {\bibfnamefont {E.}~\bibnamefont
  {Cardozo~de Oliveira}}, \bibinfo {author} {\bibfnamefont {A.}~\bibnamefont
  {Naranjo}}, \bibinfo {author} {\bibfnamefont {A.}~\bibnamefont {Pfenning}},
  \bibinfo {author} {\bibfnamefont {V.}~\bibnamefont {Lopez-Richard}}, \bibinfo
  {author} {\bibfnamefont {G.}~\bibnamefont {Marques}}, \bibinfo {author}
  {\bibfnamefont {L.}~\bibnamefont {Worschech}}, \bibinfo {author}
  {\bibfnamefont {F.}~\bibnamefont {Hartmann}}, \bibinfo {author}
  {\bibfnamefont {S.}~\bibnamefont {H\"ofling}},\ and\ \bibinfo {author}
  {\bibfnamefont {M.}~\bibnamefont {Teodoro}},\ }\bibfield  {title} {\bibinfo
  {title} {{Determination of Carrier Density and Dynamics via
  Magnetoelectroluminescence Spectroscopy in Resonant-Tunneling Diodes}},\
  }\href {https://doi.org/10.1103/PhysRevApplied.15.014042} {\bibfield
  {journal} {\bibinfo  {journal} {Phys. Rev. Applied}\ }\textbf {\bibinfo
  {volume} {15}},\ \bibinfo {pages} {014042} (\bibinfo {year}
  {2021})}\BibitemShut {NoStop}%
\bibitem [{\citenamefont {Pfenning}\ \emph {et~al.}(2017)\citenamefont
  {Pfenning}, \citenamefont {Knebl}, \citenamefont {Hartmann}, \citenamefont
  {Weih}, \citenamefont {Bader}, \citenamefont {Emmerling}, \citenamefont
  {Kamp}, \citenamefont {Höfling},\ and\ \citenamefont
  {Worschech}}]{Pfenning2017-1}%
  \BibitemOpen
  \bibfield  {author} {\bibinfo {author} {\bibfnamefont {A.}~\bibnamefont
  {Pfenning}}, \bibinfo {author} {\bibfnamefont {G.}~\bibnamefont {Knebl}},
  \bibinfo {author} {\bibfnamefont {F.}~\bibnamefont {Hartmann}}, \bibinfo
  {author} {\bibfnamefont {R.}~\bibnamefont {Weih}}, \bibinfo {author}
  {\bibfnamefont {A.}~\bibnamefont {Bader}}, \bibinfo {author} {\bibfnamefont
  {M.}~\bibnamefont {Emmerling}}, \bibinfo {author} {\bibfnamefont
  {M.}~\bibnamefont {Kamp}}, \bibinfo {author} {\bibfnamefont {S.}~\bibnamefont
  {Höfling}},\ and\ \bibinfo {author} {\bibfnamefont {L.}~\bibnamefont
  {Worschech}},\ }\bibfield  {title} {\bibinfo {title} {{Room temperature
  operation of GaSb-based resonant tunneling diodes by prewell injection}},\
  }\href {https://doi.org/10.1063/1.4973894} {\bibfield  {journal} {\bibinfo
  {journal} {Appl. Phys. Lett.}\ }\textbf {\bibinfo {volume} {110}},\ \bibinfo
  {pages} {033507} (\bibinfo {year} {2017})}\BibitemShut {NoStop}%
\bibitem [{\citenamefont {Hamamatsu}(2007)}]{Hamamatsu2007}%
  \BibitemOpen
  \bibfield  {author} {\bibinfo {author} {\bibnamefont {Hamamatsu}},\ }\href
  {https://www.hamamatsu.com/resources/pdf/etd/PMT_handbook_v3aE.pdf} {\emph
  {\bibinfo {title} {{Photomultiplier Tubes: Basics and Applications}}}},\
  \bibinfo {organization} {Hamamatsu Photonics K.K.},\ \bibinfo {address}
  {Hamamatsu},\ \bibinfo {edition} {3rd}\ ed. (\bibinfo {year}
  {2007})\BibitemShut {NoStop}%
\bibitem [{\citenamefont {PicoQuant}(2018)}]{Picoquant2018}%
  \BibitemOpen
  \bibfield  {author} {\bibinfo {author} {\bibnamefont {PicoQuant}},\
  }\href@noop {} {\emph {\bibinfo {title} {{FluoFit: Global Fluorescence Decay
  Data Analysis Software - User's Manual and Technical Data}}}},\ \bibinfo
  {organization} {PicoQuant GmbH},\ \bibinfo {address} {Berlin, Germany}
  (\bibinfo {year} {2018})\BibitemShut {NoStop}%
\bibitem [{\citenamefont {Baranowski}\ \emph {et~al.}(2012)\citenamefont
  {Baranowski}, \citenamefont {Syperek}, \citenamefont {Kudrawiec},
  \citenamefont {Misiewicz}, \citenamefont {Gupta}, \citenamefont {Wu},\ and\
  \citenamefont {Wang}}]{Baranowski2012}%
  \BibitemOpen
  \bibfield  {author} {\bibinfo {author} {\bibfnamefont {M.}~\bibnamefont
  {Baranowski}}, \bibinfo {author} {\bibfnamefont {M.}~\bibnamefont {Syperek}},
  \bibinfo {author} {\bibfnamefont {R.}~\bibnamefont {Kudrawiec}}, \bibinfo
  {author} {\bibfnamefont {J.}~\bibnamefont {Misiewicz}}, \bibinfo {author}
  {\bibfnamefont {J.~A.}\ \bibnamefont {Gupta}}, \bibinfo {author}
  {\bibfnamefont {X.}~\bibnamefont {Wu}},\ and\ \bibinfo {author}
  {\bibfnamefont {R.}~\bibnamefont {Wang}},\ }\bibfield  {title} {\bibinfo
  {title} {{Carrier dynamics in type-{II} {GaAsSb}/{GaAs} quantum wells}},\
  }\href {https://doi.org/10.1088/0953-8984/24/18/185801} {\bibfield  {journal}
  {\bibinfo  {journal} {J. Phys.: Condens. Matter}\ }\textbf {\bibinfo {volume}
  {24}},\ \bibinfo {pages} {185801} (\bibinfo {year} {2012})}\BibitemShut
  {NoStop}%
\bibitem [{\citenamefont {Romandic}\ \emph {et~al.}(2000)\citenamefont
  {Romandic}, \citenamefont {Bouwen}, \citenamefont {Goovaerts}, \citenamefont
  {Hoof},\ and\ \citenamefont {Borghs}}]{Romandic2000}%
  \BibitemOpen
  \bibfield  {author} {\bibinfo {author} {\bibfnamefont {I.}~\bibnamefont
  {Romandic}}, \bibinfo {author} {\bibfnamefont {A.}~\bibnamefont {Bouwen}},
  \bibinfo {author} {\bibfnamefont {E.}~\bibnamefont {Goovaerts}}, \bibinfo
  {author} {\bibfnamefont {C.~V.}\ \bibnamefont {Hoof}},\ and\ \bibinfo
  {author} {\bibfnamefont {G.}~\bibnamefont {Borghs}},\ }\bibfield  {title}
  {\bibinfo {title} {{Time-resolved photoluminescence spectroscopy of
  tunnelling processes in a bipolar {AlAs}/{GaAs} resonant-tunnelling
  structure}},\ }\href {https://doi.org/10.1088/0268-1242/15/7/304} {\bibfield
  {journal} {\bibinfo  {journal} {Semicond. Sci. Technol.}\ }\textbf {\bibinfo
  {volume} {15}},\ \bibinfo {pages} {665} (\bibinfo {year} {2000})}\BibitemShut
  {NoStop}%
\bibitem [{\citenamefont {Vul'}(1996)}]{Levinshtein1996}%
  \BibitemOpen
  \bibfield  {author} {\bibinfo {author} {\bibfnamefont {A.~Y.}\ \bibnamefont
  {Vul'}},\ }\bibinfo {title} {{GALLIUM ANTIMONIDE (GaSb)}},\ in\ \href
  {https://doi.org/10.1142/9789812832078_0006} {\emph {\bibinfo {booktitle}
  {Handbook Series on Semiconductor Parameters}}},\ Vol.~\bibinfo {volume} {1}\
  (\bibinfo  {publisher} {World Scientific Publishing Co. Pte. Ltd.},\ \bibinfo
  {address} {Singapore},\ \bibinfo {year} {1996})\ Chap.~\bibinfo {chapter}
  {6}, pp.\ \bibinfo {pages} {125--146}\BibitemShut {NoStop}%
\bibitem [{\citenamefont {’t Hooft}\ and\ \citenamefont {van
  Opdorp}(1983)}]{Hooft1983}%
  \BibitemOpen
  \bibfield  {author} {\bibinfo {author} {\bibfnamefont {G.~W.}\ \bibnamefont
  {’t Hooft}}\ and\ \bibinfo {author} {\bibfnamefont {C.}~\bibnamefont {van
  Opdorp}},\ }\bibfield  {title} {\bibinfo {title} {{Temperature dependence of
  interface recombination and radiative recombination in (Al, Ga)As
  heterostructures}},\ }\href {https://doi.org/10.1063/1.94105} {\bibfield
  {journal} {\bibinfo  {journal} {Appl. Phys. Lett.}\ }\textbf {\bibinfo
  {volume} {42}},\ \bibinfo {pages} {813} (\bibinfo {year} {1983})}\BibitemShut
  {NoStop}%
\bibitem [{\citenamefont {Bockelmann}(1993)}]{Bockelmann1993}%
  \BibitemOpen
  \bibfield  {author} {\bibinfo {author} {\bibfnamefont {U.}~\bibnamefont
  {Bockelmann}},\ }\bibfield  {title} {\bibinfo {title} {{Exciton relaxation
  and radiative recombination in semiconductor quantum dots}},\ }\href
  {https://doi.org/10.1103/PhysRevB.48.17637} {\bibfield  {journal} {\bibinfo
  {journal} {Phys. Rev. B}\ }\textbf {\bibinfo {volume} {48}},\ \bibinfo
  {pages} {17637} (\bibinfo {year} {1993})}\BibitemShut {NoStop}%
\bibitem [{\citenamefont {Cesar}\ \emph {et~al.}(2011)\citenamefont {Cesar},
  \citenamefont {Teodoro}, \citenamefont {Lopez-Richard}, \citenamefont
  {Marques}, \citenamefont {Jr.}, \citenamefont {Dorogan}, \citenamefont
  {Mazur},\ and\ \citenamefont {Salamo}}]{Cesar2011}%
  \BibitemOpen
  \bibfield  {author} {\bibinfo {author} {\bibfnamefont {D.~F.}\ \bibnamefont
  {Cesar}}, \bibinfo {author} {\bibfnamefont {M.~D.}\ \bibnamefont {Teodoro}},
  \bibinfo {author} {\bibfnamefont {V.}~\bibnamefont {Lopez-Richard}}, \bibinfo
  {author} {\bibfnamefont {G.~E.}\ \bibnamefont {Marques}}, \bibinfo {author}
  {\bibfnamefont {E.~M.}\ \bibnamefont {Jr.}}, \bibinfo {author} {\bibfnamefont
  {V.~G.}\ \bibnamefont {Dorogan}}, \bibinfo {author} {\bibfnamefont {Y.~I.}\
  \bibnamefont {Mazur}},\ and\ \bibinfo {author} {\bibfnamefont {G.~J.}\
  \bibnamefont {Salamo}},\ }\bibfield  {title} {\bibinfo {title} {{Carrier
  transfer in the optical recombination of quantum dots}},\ }\href
  {https://doi.org/10.1103/PhysRevB.83.195307} {\bibfield  {journal} {\bibinfo
  {journal} {Phys. Rev. B}\ }\textbf {\bibinfo {volume} {83}},\ \bibinfo
  {pages} {195307} (\bibinfo {year} {2011})}\BibitemShut {NoStop}%
\bibitem [{\citenamefont {Park}\ \emph {et~al.}(1995)\citenamefont {Park},
  \citenamefont {Chu}, \citenamefont {Han}, \citenamefont {Choi}, \citenamefont
  {Kim},\ and\ \citenamefont {Lee}}]{Park1995}%
  \BibitemOpen
  \bibfield  {author} {\bibinfo {author} {\bibfnamefont {P.~W.}\ \bibnamefont
  {Park}}, \bibinfo {author} {\bibfnamefont {H.~Y.}\ \bibnamefont {Chu}},
  \bibinfo {author} {\bibfnamefont {S.~G.}\ \bibnamefont {Han}}, \bibinfo
  {author} {\bibfnamefont {Y.~W.}\ \bibnamefont {Choi}}, \bibinfo {author}
  {\bibfnamefont {G.}~\bibnamefont {Kim}},\ and\ \bibinfo {author}
  {\bibfnamefont {E.}~\bibnamefont {Lee}},\ }\bibfield  {title} {\bibinfo
  {title} {{Optical switching mechanism based on charge accumulation effects in
  resonant tunneling diodes}},\ }\href {https://doi.org/10.1063/1.114384}
  {\bibfield  {journal} {\bibinfo  {journal} {Applied Physics Letters}\
  }\textbf {\bibinfo {volume} {67}},\ \bibinfo {pages} {1241} (\bibinfo {year}
  {1995})}\BibitemShut {NoStop}%
\bibitem [{\citenamefont {Coêlho}\ \emph {et~al.}(2004)\citenamefont
  {Coêlho}, \citenamefont {Martins-Filho}, \citenamefont {Figueiredo},\ and\
  \citenamefont {Ironside}}]{Coelho2004}%
  \BibitemOpen
  \bibfield  {author} {\bibinfo {author} {\bibfnamefont {I.~J.~S.}\
  \bibnamefont {Coêlho}}, \bibinfo {author} {\bibfnamefont {J.~F.}\
  \bibnamefont {Martins-Filho}}, \bibinfo {author} {\bibfnamefont {J.~M.~L.}\
  \bibnamefont {Figueiredo}},\ and\ \bibinfo {author} {\bibfnamefont {C.~N.}\
  \bibnamefont {Ironside}},\ }\bibfield  {title} {\bibinfo {title} {{Modeling
  of light-sensitive resonant-tunneling-diode devices}},\ }\href
  {https://doi.org/10.1063/1.1728290} {\bibfield  {journal} {\bibinfo
  {journal} {Journal of Applied Physics}\ }\textbf {\bibinfo {volume} {95}},\
  \bibinfo {pages} {8258} (\bibinfo {year} {2004})}\BibitemShut {NoStop}%
\bibitem [{\citenamefont {Pfenning}\ \emph {et~al.}(2016)\citenamefont
  {Pfenning}, \citenamefont {Hartmann}, \citenamefont {Langer}, \citenamefont
  {Kamp}, \citenamefont {Höfling},\ and\ \citenamefont
  {Worschech}}]{Pfenning2016}%
  \BibitemOpen
  \bibfield  {author} {\bibinfo {author} {\bibfnamefont {A.}~\bibnamefont
  {Pfenning}}, \bibinfo {author} {\bibfnamefont {F.}~\bibnamefont {Hartmann}},
  \bibinfo {author} {\bibfnamefont {F.}~\bibnamefont {Langer}}, \bibinfo
  {author} {\bibfnamefont {M.}~\bibnamefont {Kamp}}, \bibinfo {author}
  {\bibfnamefont {S.}~\bibnamefont {Höfling}},\ and\ \bibinfo {author}
  {\bibfnamefont {L.}~\bibnamefont {Worschech}},\ }\bibfield  {title} {\bibinfo
  {title} {"sensitivity of resonant tunneling diode photodetectors"},\ }\href
  {https://doi.org/10.1088/0957-4484/27/35/355202} {\bibfield  {journal}
  {\bibinfo  {journal} {Nanotechnology}\ }\textbf {\bibinfo {volume} {27}},\
  \bibinfo {pages} {355202} (\bibinfo {year} {2016})}\BibitemShut {NoStop}%
\bibitem [{\citenamefont {Guarin~Castro}\ \emph {et~al.}(2021)\citenamefont
  {Guarin~Castro}, \citenamefont {Pfenning}, \citenamefont {Hartmann},
  \citenamefont {Knebl}, \citenamefont {Daldin~Teodoro}, \citenamefont
  {Marques}, \citenamefont {Höfling}, \citenamefont {Bastard},\ and\
  \citenamefont {Lopez-Richard}}]{Guarin2021}%
  \BibitemOpen
  \bibfield  {author} {\bibinfo {author} {\bibfnamefont {E.~D.}\ \bibnamefont
  {Guarin~Castro}}, \bibinfo {author} {\bibfnamefont {A.}~\bibnamefont
  {Pfenning}}, \bibinfo {author} {\bibfnamefont {F.}~\bibnamefont {Hartmann}},
  \bibinfo {author} {\bibfnamefont {G.}~\bibnamefont {Knebl}}, \bibinfo
  {author} {\bibfnamefont {M.}~\bibnamefont {Daldin~Teodoro}}, \bibinfo
  {author} {\bibfnamefont {G.~E.}\ \bibnamefont {Marques}}, \bibinfo {author}
  {\bibfnamefont {S.}~\bibnamefont {Höfling}}, \bibinfo {author}
  {\bibfnamefont {G.}~\bibnamefont {Bastard}},\ and\ \bibinfo {author}
  {\bibfnamefont {V.}~\bibnamefont {Lopez-Richard}},\ }\bibfield  {title}
  {\bibinfo {title} {Optical mapping of nonequilibrium charge carriers},\
  }\href {https://doi.org/10.1021/acs.jpcc.1c02173} {\bibfield  {journal}
  {\bibinfo  {journal} {The Journal of Physical Chemistry C}\ }\textbf
  {\bibinfo {volume} {125}},\ \bibinfo {pages} {14741} (\bibinfo {year}
  {2021})}\BibitemShut {NoStop}%
\bibitem [{\citenamefont {Pfenning}\ \emph {et~al.}(2018)\citenamefont
  {Pfenning}, \citenamefont {Hartmann}, \citenamefont {Weih}, \citenamefont
  {Emmerling}, \citenamefont {Worschech},\ and\ \citenamefont
  {Höfling}}]{Pfenning2018}%
  \BibitemOpen
  \bibfield  {author} {\bibinfo {author} {\bibfnamefont {A.}~\bibnamefont
  {Pfenning}}, \bibinfo {author} {\bibfnamefont {F.}~\bibnamefont {Hartmann}},
  \bibinfo {author} {\bibfnamefont {R.}~\bibnamefont {Weih}}, \bibinfo {author}
  {\bibfnamefont {M.}~\bibnamefont {Emmerling}}, \bibinfo {author}
  {\bibfnamefont {L.}~\bibnamefont {Worschech}},\ and\ \bibinfo {author}
  {\bibfnamefont {S.}~\bibnamefont {Höfling}},\ }\bibfield  {title} {\bibinfo
  {title} {p-type doped alassb/gasb resonant tunneling diode photodetector for
  the mid-infrared spectral region},\ }\href
  {https://doi.org/https://doi.org/10.1002/adom.201800972} {\bibfield
  {journal} {\bibinfo  {journal} {Advanced Optical Materials}\ }\textbf
  {\bibinfo {volume} {6}},\ \bibinfo {pages} {1800972} (\bibinfo {year}
  {2018})}\BibitemShut {NoStop}%
\bibitem [{\citenamefont {Bastard}\ \emph {et~al.}(1991)\citenamefont
  {Bastard}, \citenamefont {Brum},\ and\ \citenamefont
  {Ferreira}}]{Bastard1991}%
  \BibitemOpen
  \bibfield  {author} {\bibinfo {author} {\bibfnamefont {G.}~\bibnamefont
  {Bastard}}, \bibinfo {author} {\bibfnamefont {J.}~\bibnamefont {Brum}},\ and\
  \bibinfo {author} {\bibfnamefont {R.}~\bibnamefont {Ferreira}},\ }\bibfield
  {title} {\bibinfo {title} {{Electronic States in Semiconductor
  Heterostructures}},\ }in\ \href
  {https://doi.org/https://doi.org/10.1016/S0081-1947(08)60092-2} {\emph
  {\bibinfo {booktitle} {{Semiconductor Heterostructures and
  Nanostructures}}}},\ \bibinfo {series} {Solid State Physics}, Vol.~\bibinfo
  {volume} {44},\ \bibinfo {editor} {edited by\ \bibinfo {editor}
  {\bibfnamefont {H.}~\bibnamefont {Ehrenreich}}\ and\ \bibinfo {editor}
  {\bibfnamefont {D.}~\bibnamefont {Turnbull}}}\ (\bibinfo  {publisher}
  {Academic Press},\ \bibinfo {year} {1991})\ pp.\ \bibinfo {pages}
  {229--415}\BibitemShut {NoStop}%
\bibitem [{\citenamefont {Larkin}\ \emph {et~al.}(2009)\citenamefont {Larkin},
  \citenamefont {Ujevic},\ and\ \citenamefont {Avrutin}}]{Larkin2009}%
  \BibitemOpen
  \bibfield  {author} {\bibinfo {author} {\bibfnamefont {I.~A.}\ \bibnamefont
  {Larkin}}, \bibinfo {author} {\bibfnamefont {S.}~\bibnamefont {Ujevic}},\
  and\ \bibinfo {author} {\bibfnamefont {E.~A.}\ \bibnamefont {Avrutin}},\
  }\bibfield  {title} {\bibinfo {title} {{Tunneling escape time from a
  semiconductor quantum well in an electric field}},\ }\bibfield  {journal}
  {\bibinfo  {journal} {J. Appl. Phys.}\ }\textbf {\bibinfo {volume} {106}},\
  \href {https://doi.org/10.1063/1.3259414} {10.1063/1.3259414} (\bibinfo
  {year} {2009}),\ \bibinfo {note} {113701}\BibitemShut {NoStop}%
\bibitem [{\citenamefont {Bastard}(1990)}]{bastard1990}%
  \BibitemOpen
  \bibfield  {author} {\bibinfo {author} {\bibfnamefont {G.}~\bibnamefont
  {Bastard}},\ }\href@noop {} {\emph {\bibinfo {title} {{Wave Mechanics Applied
  to Semiconductor Heterostructures}}}}\ (\bibinfo  {publisher} {les éditions
  de physique},\ \bibinfo {address} {Paris},\ \bibinfo {year}
  {1990})\BibitemShut {NoStop}%
\bibitem [{\citenamefont {Guarin~Castro}(2021)}]{Guarin2021_phd}%
  \BibitemOpen
  \bibfield  {author} {\bibinfo {author} {\bibfnamefont {E.~D.}\ \bibnamefont
  {Guarin~Castro}},\ }\emph {\bibinfo {title} {{Charge Carrier Dynamics and
  Optoelectronic Properties in Quantum Tunneling Heterostructures}}},\ \href
  {https://repositorio.ufscar.br/handle/ufscar/14972} {Ph.D. thesis},\ \bibinfo
   {school} {Federal University of Sao Carlos, Programa de Pos-graduacao em
  Fisica}, \bibinfo {address} {Sao Carlos, SP} (\bibinfo {year}
  {2021})\BibitemShut {NoStop}%
\bibitem [{\citenamefont {Manasreh}(2005)}]{Manasreh2005}%
  \BibitemOpen
  \bibfield  {author} {\bibinfo {author} {\bibfnamefont {O.}~\bibnamefont
  {Manasreh}},\ }\href@noop {} {\emph {\bibinfo {title} {Semiconductor
  Heterojunctions and Nanostructures}}}\ (\bibinfo  {publisher} {McGraw-Hill,
  Inc.},\ \bibinfo {year} {2005})\BibitemShut {NoStop}%
\bibitem [{\citenamefont {Madelung}(2004)}]{Madelung2004}%
  \BibitemOpen
  \bibfield  {author} {\bibinfo {author} {\bibfnamefont {O.}~\bibnamefont
  {Madelung}},\ }\href@noop {} {\emph {\bibinfo {title} {{Semiconductors: Data
  Handbook}}}},\ \bibinfo {edition} {3rd}\ ed.\ (\bibinfo  {publisher}
  {Springer},\ \bibinfo {address} {Berlin},\ \bibinfo {year}
  {2004})\BibitemShut {NoStop}%
\bibitem [{\citenamefont {Vurgaftman}\ \emph {et~al.}(2001)\citenamefont
  {Vurgaftman}, \citenamefont {Meyer},\ and\ \citenamefont
  {Ram-Mohan}}]{Vurgaftman2001}%
  \BibitemOpen
  \bibfield  {author} {\bibinfo {author} {\bibfnamefont {I.}~\bibnamefont
  {Vurgaftman}}, \bibinfo {author} {\bibfnamefont {J.~R.}\ \bibnamefont
  {Meyer}},\ and\ \bibinfo {author} {\bibfnamefont {L.~R.}\ \bibnamefont
  {Ram-Mohan}},\ }\bibfield  {title} {\bibinfo {title} {{Band parameters for
  III–V compound semiconductors and their alloys}},\ }\href
  {https://doi.org/10.1063/1.1368156} {\bibfield  {journal} {\bibinfo
  {journal} {J. Appl. Phys.}\ }\textbf {\bibinfo {volume} {89}},\ \bibinfo
  {pages} {5815} (\bibinfo {year} {2001})}\BibitemShut {NoStop}%
\bibitem [{\citenamefont {Van~Hoof}\ \emph {et~al.}(1992)\citenamefont
  {Van~Hoof}, \citenamefont {Goovaerts},\ and\ \citenamefont
  {Borghs}}]{VanHoof1992_TRPL}%
  \BibitemOpen
  \bibfield  {author} {\bibinfo {author} {\bibfnamefont {C.}~\bibnamefont
  {Van~Hoof}}, \bibinfo {author} {\bibfnamefont {E.}~\bibnamefont
  {Goovaerts}},\ and\ \bibinfo {author} {\bibfnamefont {G.}~\bibnamefont
  {Borghs}},\ }\bibfield  {title} {\bibinfo {title} {{Sequential hole tunneling
  in n-type AlAs/GaAs resonant-tunneling structures from time-resolved
  photoluminescence}},\ }\href {https://doi.org/10.1103/PhysRevB.46.6982}
  {\bibfield  {journal} {\bibinfo  {journal} {Phys. Rev. B}\ }\textbf {\bibinfo
  {volume} {46}},\ \bibinfo {pages} {6982} (\bibinfo {year}
  {1992})}\BibitemShut {NoStop}%
\bibitem [{\citenamefont {Cardozo~de Oliveira}\ \emph
  {et~al.}(2018)\citenamefont {Cardozo~de Oliveira}, \citenamefont {Pfenning},
  \citenamefont {Guarin~Castro}, \citenamefont {Teodoro}, \citenamefont {dos
  Santos}, \citenamefont {Lopez-Richard}, \citenamefont {Marques},
  \citenamefont {Worschech}, \citenamefont {Hartmann},\ and\ \citenamefont
  {H\"ofling}}]{Edson}%
  \BibitemOpen
  \bibfield  {author} {\bibinfo {author} {\bibfnamefont {E.~R.}\ \bibnamefont
  {Cardozo~de Oliveira}}, \bibinfo {author} {\bibfnamefont {A.}~\bibnamefont
  {Pfenning}}, \bibinfo {author} {\bibfnamefont {E.~D.}\ \bibnamefont
  {Guarin~Castro}}, \bibinfo {author} {\bibfnamefont {M.~D.}\ \bibnamefont
  {Teodoro}}, \bibinfo {author} {\bibfnamefont {E.~C.}\ \bibnamefont {dos
  Santos}}, \bibinfo {author} {\bibfnamefont {V.}~\bibnamefont
  {Lopez-Richard}}, \bibinfo {author} {\bibfnamefont {G.~E.}\ \bibnamefont
  {Marques}}, \bibinfo {author} {\bibfnamefont {L.}~\bibnamefont {Worschech}},
  \bibinfo {author} {\bibfnamefont {F.}~\bibnamefont {Hartmann}},\ and\
  \bibinfo {author} {\bibfnamefont {S.}~\bibnamefont {H\"ofling}},\ }\bibfield
  {title} {\bibinfo {title} {{Electroluminescence on-off ratio control of
  $\mathit{n}\mathit{\text{\ensuremath{-}}}\mathit{i}\mathit{\text{\ensuremath{-}}}\mathit{n}$
  GaAs/AlGaAs-based resonant tunneling structures}},\ }\href
  {https://doi.org/10.1103/PhysRevB.98.075302} {\bibfield  {journal} {\bibinfo
  {journal} {Phys. Rev. B}\ }\textbf {\bibinfo {volume} {98}},\ \bibinfo
  {pages} {075302} (\bibinfo {year} {2018})}\BibitemShut {NoStop}%
\bibitem [{\citenamefont {Snow}\ \emph {et~al.}(1989)\citenamefont {Snow},
  \citenamefont {Westland}, \citenamefont {Ryan}, \citenamefont {Kerr},
  \citenamefont {Munekata},\ and\ \citenamefont {Chang}}]{Snow1989}%
  \BibitemOpen
  \bibfield  {author} {\bibinfo {author} {\bibfnamefont {P.}~\bibnamefont
  {Snow}}, \bibinfo {author} {\bibfnamefont {D.}~\bibnamefont {Westland}},
  \bibinfo {author} {\bibfnamefont {J.}~\bibnamefont {Ryan}}, \bibinfo {author}
  {\bibfnamefont {T.}~\bibnamefont {Kerr}}, \bibinfo {author} {\bibfnamefont
  {H.}~\bibnamefont {Munekata}},\ and\ \bibinfo {author} {\bibfnamefont
  {L.}~\bibnamefont {Chang}},\ }\bibfield  {title} {\bibinfo {title} {{Hot
  carrier cooling in GaSb: Bulk and quantum wells}},\ }\href
  {https://doi.org/https://doi.org/10.1016/0749-6036(89)90393-5} {\bibfield
  {journal} {\bibinfo  {journal} {Superlattices Microstruct.}\ }\textbf
  {\bibinfo {volume} {5}},\ \bibinfo {pages} {595} (\bibinfo {year}
  {1989})}\BibitemShut {NoStop}%
\bibitem [{\citenamefont {Pelouch}\ and\ \citenamefont
  {Schlie}(1995)}]{Pelouch1995}%
  \BibitemOpen
  \bibfield  {author} {\bibinfo {author} {\bibfnamefont {W.~S.}\ \bibnamefont
  {Pelouch}}\ and\ \bibinfo {author} {\bibfnamefont {L.~A.}\ \bibnamefont
  {Schlie}},\ }\bibfield  {title} {\bibinfo {title} {{Ultrafast carrier
  dynamics in GaSb}},\ }\href {https://doi.org/10.1063/1.114153} {\bibfield
  {journal} {\bibinfo  {journal} {Appl. Phys. Lett.}\ }\textbf {\bibinfo
  {volume} {66}},\ \bibinfo {pages} {82} (\bibinfo {year} {1995})}\BibitemShut
  {NoStop}%
\bibitem [{\citenamefont {Pellemans}\ \emph {et~al.}(1996)\citenamefont
  {Pellemans}, \citenamefont {Wenckebach},\ and\ \citenamefont
  {Planken}}]{Pellemans1996}%
  \BibitemOpen
  \bibfield  {author} {\bibinfo {author} {\bibfnamefont {H.~P.~M.}\
  \bibnamefont {Pellemans}}, \bibinfo {author} {\bibfnamefont {W.~T.}\
  \bibnamefont {Wenckebach}},\ and\ \bibinfo {author} {\bibfnamefont
  {P.~C.~M.}\ \bibnamefont {Planken}},\ }\bibfield  {title} {\bibinfo {title}
  {{Sub-Picosecond Far-Infrared Transient-Grating Measurements of Electron
  Cooling in InAs and GaSb}},\ }in\ \href
  {https://link.springer.com/chapter/10.1007/978-3-642-80314-7_174} {\emph
  {\bibinfo {booktitle} {{Ultrafast Phenomena X}}}},\ \bibinfo {editor} {edited
  by\ \bibinfo {editor} {\bibfnamefont {P.~F.}\ \bibnamefont {Barbara}},
  \bibinfo {editor} {\bibfnamefont {J.~G.}\ \bibnamefont {Fujimoto}}, \bibinfo
  {editor} {\bibfnamefont {W.~H.}\ \bibnamefont {Knox}},\ and\ \bibinfo
  {editor} {\bibfnamefont {W.}~\bibnamefont {Zinth}}}\ (\bibinfo  {publisher}
  {Springer Berlin Heidelberg},\ \bibinfo {address} {Berlin, Heidelberg},\
  \bibinfo {year} {1996})\ pp.\ \bibinfo {pages} {400--401}\BibitemShut
  {NoStop}%
\bibitem [{\citenamefont {Smith}\ \emph {et~al.}(1998)\citenamefont {Smith},
  \citenamefont {O'Sullivan}, \citenamefont {Rota}, \citenamefont {Maciel},\
  and\ \citenamefont {Ryan}}]{Smith1998}%
  \BibitemOpen
  \bibfield  {author} {\bibinfo {author} {\bibfnamefont {D.}~\bibnamefont
  {Smith}}, \bibinfo {author} {\bibfnamefont {E.}~\bibnamefont {O'Sullivan}},
  \bibinfo {author} {\bibfnamefont {L.}~\bibnamefont {Rota}}, \bibinfo {author}
  {\bibfnamefont {A.}~\bibnamefont {Maciel}},\ and\ \bibinfo {author}
  {\bibfnamefont {J.}~\bibnamefont {Ryan}},\ }\bibfield  {title} {\bibinfo
  {title} {{Ultrafast optical response and inter-valley scattering in GaSb/AlSb
  quantum wells}},\ }\href
  {https://doi.org/https://doi.org/10.1016/S1386-9477(98)00034-4} {\bibfield
  {journal} {\bibinfo  {journal} {Phys. E}\ }\textbf {\bibinfo {volume} {2}},\
  \bibinfo {pages} {156} (\bibinfo {year} {1998})}\BibitemShut {NoStop}%
\end{thebibliography}%

\end{document}